\documentclass{aa}  
\usepackage{graphicx}
\usepackage{longtable}
\usepackage{txfonts}
\begin{document}
\def\ang{$\rm\AA$}

   \title{Quantitative optical and near-infrared spectroscopy of
H$_2$ towards HH91A}

   \author{R. Gredel
          \inst{1}
          \fnmsep\thanks{
          Based on observations collected at the Centro Astronomico Hispano
Aleman (CAHA) Calar Alto, operated jointly by the Max-Planck Institute for Astronomy
and the Instituto de Astrofisica de Andalucia (CSIC). Based on observations collected
at the European Southern Observatory, La Silla, Chile.}
          }

   \offprints{R. Gredel}

   \institute{Max-Planck Institute for Astronomy, K\"onigstuhl 17, 
             D-69117 Heidelberg, Germany\\
              \email{gredel@mpia.de}
             }

   \date{Received July 13, 2007; accepted August 31, 2007}

\abstract
%contents
{}
%aims
{Optical and near-infrared spectroscopy of molecular hydrogen in interstellar shocks provide 
a very powerful probe to the physical conditions that prevail in interstellar shocks.}
%method
{Integral-field spectroscopy of H$_2$ in the optical wavelength region
and complementary long-slit near-infrared spectroscopy
towards \object{HH91A} is used to characterize the ro-vibrational population distribution 
among H$_2$ levels with excitation energies up to 30\,000 cm$^{-1}$. 
}
%results
{The detection of some 200 ro-vibrational lines of molecular hydrogen ranging between
7700 \ang\ and 2.3 $\mu$m is reported.  Emission lines which arise 
from vibrational levels up to $v'=8$
are detected.  
The H$_2$ emission arises from thermally excited gas where the bulk
of the material is at a temperature of 2750~K and where 1\% is at 6000~K. 
The total column density of shocked molecular hydrogen is
$N$(H$_2) = 10^{18}$ cm$^{-2}$. Non-thermal excitation scenarios such as
UV fluorescence do not contribute to the H$_2$ excitation observed towards HH91A.
}
%conclusions
{The emission of molecular hydrogen towards HH91A is explained in terms of a slow J-shock 
which propagates into a low-density medium which has been swept-up by previous episodes
of outflows which have occurred in the evolved \object{HH90}/\object{91} complex.
} 

   \maketitle
%
%________________________________________________________________

\section{Introduction}

The study of molecular hydrogen emission lines in star-forming regions
provides a powerful tool to gain insight into the physical processes which occur during
the early stages of star formation. Outflows from young stellar objects drive powerful 
shock waves into the interstellar medium. The heating associated with the shocks 
can give rise to the excitation and dissociation of H$_2$. 
For low-mass protostars, the total H$_2$ luminosities are proportional
to the accretion rates during the early phases of the protostellar evolution, 
and evidence exists that the proportionality extents to the high-mass stellar
regime as well (Froebrich et al. \cite{froebrich}, Davis et al. 
\cite{davis}, Caratti o Garatti et al. 2006, Gredel \cite{gredel06}). 
These findings support a scenario where 
high-mass star formation proceeds via accretion as well but at significantly
larger accretion rates, compared to their low-mass counterparts
(eg. McKee \& Tan \cite{mckee}, Yorke \& Sonnenhalter \cite{yorke}). 
The shock waves that lead to H$_2$ emission 
are either continuous (C-shock) or jump type (J-shock), depending on the
physical conditions in the pre-shock gas, such as the magnetic field strength and
the degree of ionization. The physical parameters and the H$_2$ luminosities
depend on the evolutionary state
of the driving source. For instance, jets from Class 0 sources travel in the high density
gas from which the protostars are forming, and strong H$_2$ emission from C-type shocks
is expected. Jets from older protostars propagate
into a medium at lower density, since the mass loss during the early phase of the
protostellar evolution has already swept up much of the ambient gas - conditions which 
favor dissociative J-type shocks (Caratti o Garatti et al 2006). 
The C-type shocks produce a large column of warm gas in
the v'=0 levels of H$_2$, while the J-type shocks produce large columns of hot
gas of several 1000~K in the higher vibrational levels (Cabrit et al \cite{cabrit},
Smith, O'Connell \& Davis (2007), and references therein).

Comprehensive near-infrared spectroscopy of molecular hydrogen emission in  
Herbig-Haro (HH) objects covering the J-, H-, and Ks-bands 
have consequently been used to probe the physical conditions in molecular outflows from
protostars (eg. Caratti o Garatti et al.  \cite{caratti}, and references therein).
The H$_2$ emission in Herbig-Haro objects is in general dominated by thermal emission
which arise from rotational levels in v'=1--5.  
In general, J-type shocks are preferred to explain the observed H$_2$ emission
(Smith 1994; Gredel 1994, 1996; McCoey et al. \cite{mccoey}; Nisini et al. \cite{nisini}, 
among others), yet it has been noted that the population distribution among
ro-vibrational levels in v'=1--5 is not the best discriminator to unambiguously infer 
the type of shock that is at work (Flower et al. \cite{flower}).
Using emission from pure rotational lines in the (0,0) band of H$_2$, 
Giannini et al. (2006) 
convincingly demonstrated that the emission towards HH54 
arises from a steady-state J-shock. 

In the following sections, a quantitative study of the molecular hydrogen
emission in HH91A is presented. The novel aspect of the present study 
is given by the study of H$_2$ emission lines in the optical
wavelength region between 7700--8700 \ang, and the study of relatively
faint emission lines which arise from very high-excitation  ro-vibrational levels in
the near-infrared.
HH91A is part of the HH90/91 complex of Herbig-Haro (HH) objects which is located in the 
L1630 cloud.  A comprehensive optical/infrared/millimeter study has been performed by 
Gredel, Reipurth \& Heathcote (\cite{gredel92}). Complementary near-infrared observations  
were presented by Davis et al. (\cite{davis94}).
The complex shows widespread and diffuse emission of molecular hydrogen which 
extends over several square arcmin.  Superimposed are a number of very bright H$_2$ knots
such as HH91A. 
The bulk of the H$_2$ emission from HH91A arises from hot gas at a temperature of 
2750~K (Gredel et al. 1992). Deep near-infrared imaging by Moneti \& Reipurth (1995)
did not detect the energy source that drives the HH90/91 outflow. 
HH90/91 has been characterized to present a fairly evolved state of
Herbig-Haro objects (Gredel et al. 1992).  

Because the H$_2$ emission towards HH91A is very bright indeed, HH91A affords the
possibility to study emission from very high-excitation levels of H$_2$ which
arise in the optical and near-infrared wavelength regions. 
The population density in these levels provides a very sensitive discriminator 
among the various physical processes that contribute to the excitation of H$_2$ in
shocks. The optical spectra of HH91A obtained with the integral field spectrograph 
PMAS at the Calar Alto 3.5m telescope are described in Sect.~\ref{obs}, together 
with complementary near-infrared spectra obtained with SOFI at the ESO/La Silla New
Technology Telescope. The
results are summarized in Sect~\ref{results}, which also contains a description of
a theoretical model of H$_2$ which is compared with the observations. We conclude
with a discussion of the significance of non-thermal excitation scenarios of the 
H$_2$ emission towards HH91A in Sect.~\ref{discussion}.

\section{Observations and Reduction}
\label{obs}

Optical spectroscopy of HH91A was carried during the nights of Feb 15 and 16, 2004, using
the Potsdam Multi-Aperture Spectrophotometer
PMAS at the Calar Alto 3.5m telescope (Roth et al. \cite{roth}). 
PMAS is an integral field instrument and was
used in its standard configuration with a 16 x 16 lenslet
array of $8'' \times 8''$ on the sky. The R1200 reflective grating provided a spectral 
resolution of approximately R = $\lambda/\Delta \lambda = 10\,000$. The grating was
used at encoder settings of 49$^\circ$ and 46$^\circ$, which resulted in a spectral coverage of 
7690--8270 \ang\ and 8400--8980 \ang,
respectively. Sky-subtraction was achieved using the nod-and-shuffle technique 
(Roth et al. \cite{roth2}), where the charge-shuffle mode of the CCDs is used to perform 
beam-switching between HH91A and a sky position 
during ongoing integrations. This mode results in a very high degree in the accuracy 
of the sky subtraction. Atmospheric transmission was corrected via the observation of various
telluric standard stars. The data were obtained during non-photometric observing conditions which
did not allow to derive a flux calibration. However, a number of
H$_2$ emission lines in the 7700--8700 \ang\ optical spectra arise
from the same upper ro-vibrational levels than emission lines in the near-infrared wavelength
region.  Examples are the (4,1) S(3) line near 8500 \ang\ and the (4,2) S(3) and (4,2) Q(5)
lines in the J-band which arise from v'= 4, J'=5, or the (3,0) S(3), (3,0) Q(5) in
the optical and the (3,1) S(3) and (3,1) Q(5) lines, which arise from v'=3, J'=5. 
The near-infrared observations were obtained during photometric conditions, 
and the H$_2$ population densities inferred from the near-infrared observations were
used to obtain a relative flux calibration for the optical spectrum. 
We ignore reddening towards HH91A following Gredel et al. (1992).
The PMAS observations afford the possibility to study spatial variations in line ratios
across the H$_2$ line emitting regions. The H$_2$ emission detected in the optical 
wavelength regime is very faint indeed. 
The spectra in a PMAS pixel (0.25 square arcseconds) have a
too low signal to noise ratio to derive meaningful conclusions. We thus 
proceed and sum up all spectra over the central $5''$ emission region 
and surrender on the potential of the PMAS observations to study changes in the 
H$_2$ excitation across HH91A. 

The near-infrared spectra cover the J, H, and Ks-band atmospheric windows and were obtained during
the nights of Dec 20 and 21, 2003, using SOFI at the La Silla New Technology Telescope NTT. 
The reduction of the data and the flux calibration follows recipes described elsewhere 
(e.g. Gredel \cite{gredel06}). The observations were carried out during photometric conditions and
thus allow to infer total column densities in the various ro-vibrational levels of H$_2$
(see Gredel \cite{gredel06} for details). The spectra were obtained using a slit width
of $0\farcs6$. The blue grism GB in order 1 and the HR grism in orders 2 and 1 were used which 
provide spectral resolutions of R = 600, 1560, and 1800 in the J-, H-, and Ks-bands, respectively.
The one-dimensional spectra were extracted using a 20-pixel
extraction window along the slit ($5\farcs8$),  which corresponds to an 'aperture' 
of 3.6 square arcseconds or a solid angle of $\Omega = 9 \times 10^{-11}$ sr$^{-1}$.

%
%__________________________________________________ One column table
   \begin{table}
      \caption[]{Optical emission lines of H$_2$ detected with PMAS }
         \label{tableone}
     $$ 
         \begin{array}{p{0.25\linewidth}lrrr}
         \hline
         \noalign{\smallskip}
  Line &  \mathrm{Wavelength} & \mathrm{Flux}F & N\mathrm(v'J') \\
                                  &  (\ang)       & (10^{-19}\mathrm{Wm}^{-2})
                                                & (10^{14} \mathrm{cm}^{-2})\\
            \noalign{\smallskip}
            \hline
            \noalign{\smallskip}

( 3, 0)S( 8)  & 7781.  &    2.6(  0.5 ) &  1.3 (  0.3 )\\ 
( 3, 0)S( 7)  & 7784.  &    8.1(  0.8 ) &  3.9 (  0.4 )\\ 
( 7, 3)S( 9)  & 7782.  &    \le 1 &  \le 0.1\\ 
( 3, 0)S( 9)  & 7793.  &    6.9(  0.7 ) &  3.2 (  0.6 )\\
( 3, 0)S( 6)  & 7804.  &    2.9(  0.3 ) &  1.6 (  0.3 )\\ 
( 3, 0)S(10)  & 7821.  &    \le 0.4 &  \le 0.2 \\ 
( 3, 0)S( 5)  & 7840.  &    8.1(  0.8 ) &  4.9 (  1.0 )\\ 
( 3, 0)S(11)  & 7865.  &    3.0(  0.3 ) &  1.5 (  0.3 )\\ 
( 3, 0)S( 4)  & 7892.  &    1.5(  0.3 ) &  1.1 (  0.3 )\\ 
( 3, 0)S(12)  & 7924.  &    \le 0.4 &  \le 0.2 \\ 
( 3, 0)S( 3)  & 7962.  &    5.8(  0.6 ) &  4.9 (  1.0 )\\ 
( 7, 3)S(11)  & 7978.  &    \le 0.4 &  \le 0.1\\ 
( 3, 0)S(13)  & 7999.  &    0.9(  0.3 ) &  0.6 (  0.2 )\\ 
( 3, 0)S( 2)  & 8049.  &    1.0(  0.3 ) &  1.1 (  0.4 )\\ 
( 3, 0)S(14)  & 8090.  &    \le 0.4 &  \le 0.3 \\ 
( 3, 0)S( 1)  & 8153.  &    2.3(  0.3 ) &  3.6 (  0.7 )\\ 
( 4, 1)S( 4)  & 8390.  &    2.8(  0.3 ) &  0.7 (  0.1 )\\ 
( 4, 1)S(11)  & 8398.  &    4.1(  0.4 ) &  0.8 (  0.2 )\\ 
( 4, 1)S( 3)  & 8462.  &    5.9(  0.6 ) &  1.6 (  0.3 )\\ 
( 4, 1)S(12)  & 8471.  &    1.0(  0.3 ) &  0.2 (  0.0 )\\ 
( 8, 4)S( 9)  & 8496.  &    \le 0.4 &  \le 0.2\\ 
( 3, 0)Q( 1)  & 8500.  &    1.2(  0.3 ) &  2.8 (  0.7 )\\ 
( 3, 0)Q( 2)  & 8525.  &    \le 0.4 &  \le 1.3 \\ 
( 4, 1)S( 2)  & 8552.  &    1.4(  0.3 ) &  0.5 (  0.1 )\\ 
( 3, 0)Q( 3)  & 8563.  &    3.3(  0.3 ) & 11.4 (  2.3 )\\ 
( 4, 1)S(13)  & 8562.  &    3.3(  0.3 ) &  0.8 (  0.2 )\\ 
( 4, 1)S( 1)  & 8662.  &    2.5(  0.3 ) &  1.3 (  0.3 )\\ 
( 4, 1)S(14)  & 8673.  &    1.5(  0.3 ) &  0.4 (  0.1 )\\ 
( 3, 0)Q( 5)  & 8677.  &    1.5(  0.3 ) &  5.4 (  1.1 )\\ 
( 3, 0)Q( 4)  & 8613.  &    \le 0.4 &  \le 1.4 \\ 
( 3, 0)O( 2)  & 8750.  &    \le 0.6 &  \le 1.2 \\ 
( 3, 0)Q( 6)  & 8753.  &    \le 0.6 &  \le 2.3 \\ 
( 8, 4)S(11)  & 8763.  &    \le 0.6 &  \le 0.1 \\ 
( 4, 1)S( 0)  & 8792.  &    \le 0.6 &  \le 0.5 \\ 
( 4, 1)S(15)  & 8803.  &    \le 0.6 &  \le 0.2 \\ 
            \noalign{\smallskip}
            \hline
         \end{array}
     $$ 
%\begin{list}{}{}
%\item[$^{\mathrm{a}}$] This is footnote a
%\end{list}
   \end{table}

\section{Results}
\label{results}

\subsection{Optical spectroscopy using PMAS}
\label{optical}

\begin{figure*}
   \centering
   \includegraphics[angle=-90,width=20cm]{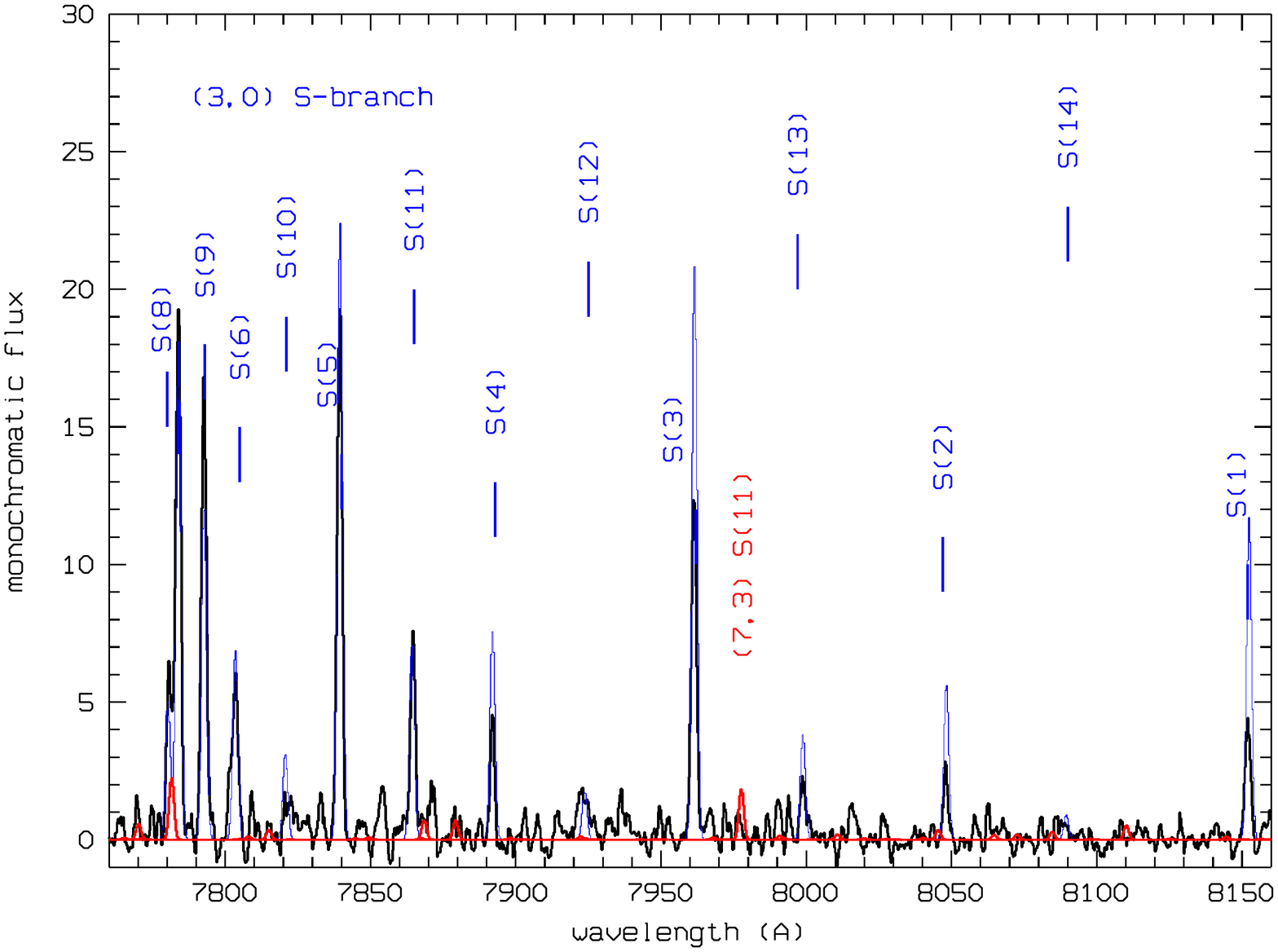}
   \caption{
Observed spectrum towards HH91A obtained with PMAS, with monochromatic 
fluxes plotted versus wavelength (in \ang). The position of the (3,0) S(1)--S(14) 
lines are indicated. The (7,3) S(11) line near 7978 \ang is marginally detected. 
A model spectrum with emission from vibrational levels $v'$=3 and 7 is
color-coded in blue and red, respectively (cf. Sect.~\ref{discussion}). 
The emission near 8727 \ang arises from atomic carbon.
               }
              \label{red1}
    \end{figure*}

\begin{figure*}
   \centering
   \includegraphics[angle=-90,width=20cm]{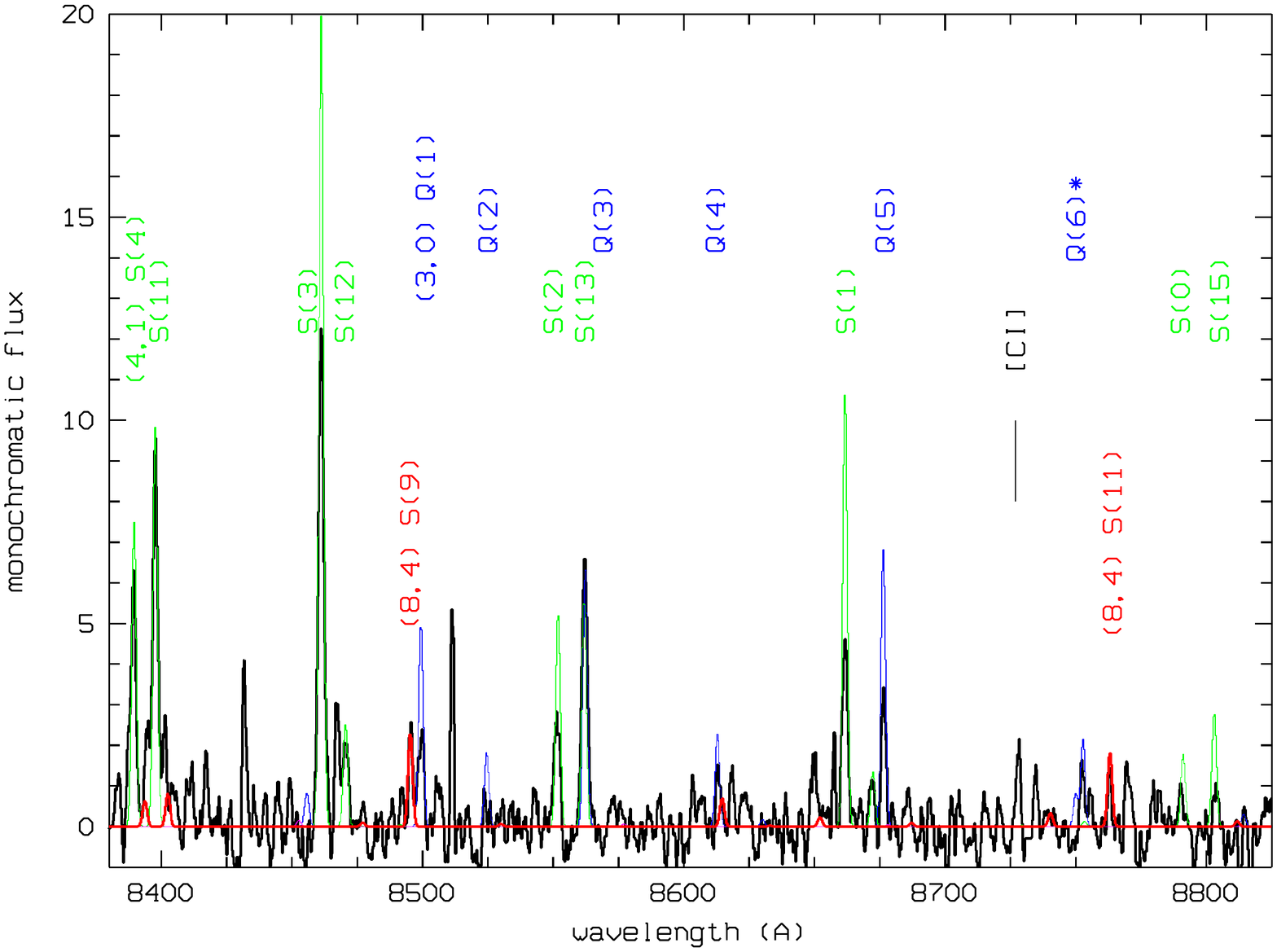}
   \caption{
Observed spectrum towards HH91A obtained with PMAS, with monochromatic 
fluxes (in relative units) plotted versus wavelength (in \ang). The position of various
emission lines in the (3,0) and (4,1) bands are indicated.
The (8,4) S(9) and (8,4) S(11) lines near 8500 \ang\ and 8685 \ang,
respectively, are marginally detected. 
The emission which is color-coded in blue, green, and red 
corresponds to a model spectrum with
emission from vibrational levels $v'$=3, 4, and 8, respectively
(cf. Sect.~\ref{discussion}).
               }
              \label{red2}
    \end{figure*}

The optical spectra obtained towards HH91A 
are shown in Figs.~\ref{red1}--\ref{red2}. Emission from the (3,0) S(1)--S(14) lines is
detected long-ward of the (3,0) S-branch band head marked by the (3,0) S(8) line
at 7781 \ang\ (Fig.~\ref{red1}).  The (3,0) Q(1)--Q(6) lines are
detected as well, together with several lines in the (4,1) S-branch
 (Fig.~\ref{red2}).  The (8,4) S(9) line near 8496 \ang\ is clearly detected.
A wavelet analysis of the optical spectra
confirms the marginal detection of the (7,3) S(11) line at 7978 \ang, and of the 
(8,4) S(11) lines near 8763 \ang. The signal to noise ratio in the
latter two lines is very low indeed, and as a standalone result, the claim of the
detection of the latter three emission lines in our spectra may be disputed.
The red lines in Figs.~1 and 2
reproduces the expected fluxes in the (7,3) S(11) and (8,4) S(9) and S(11) lines
from a model which is presented in detail below. The model is based on the analysis
of the full
set of some 200 observed H$_2$ emission lines towards HH91A and substantiates
the result from the wavelet analysis, which indicates that emission from the 
(7,3) and (8,4) bands towards HH91A is detected.

\subsection{Near-infrared spectroscopy using SOFI}

\begin{figure*}
   \centering
   \includegraphics[angle=-90,width=20cm]{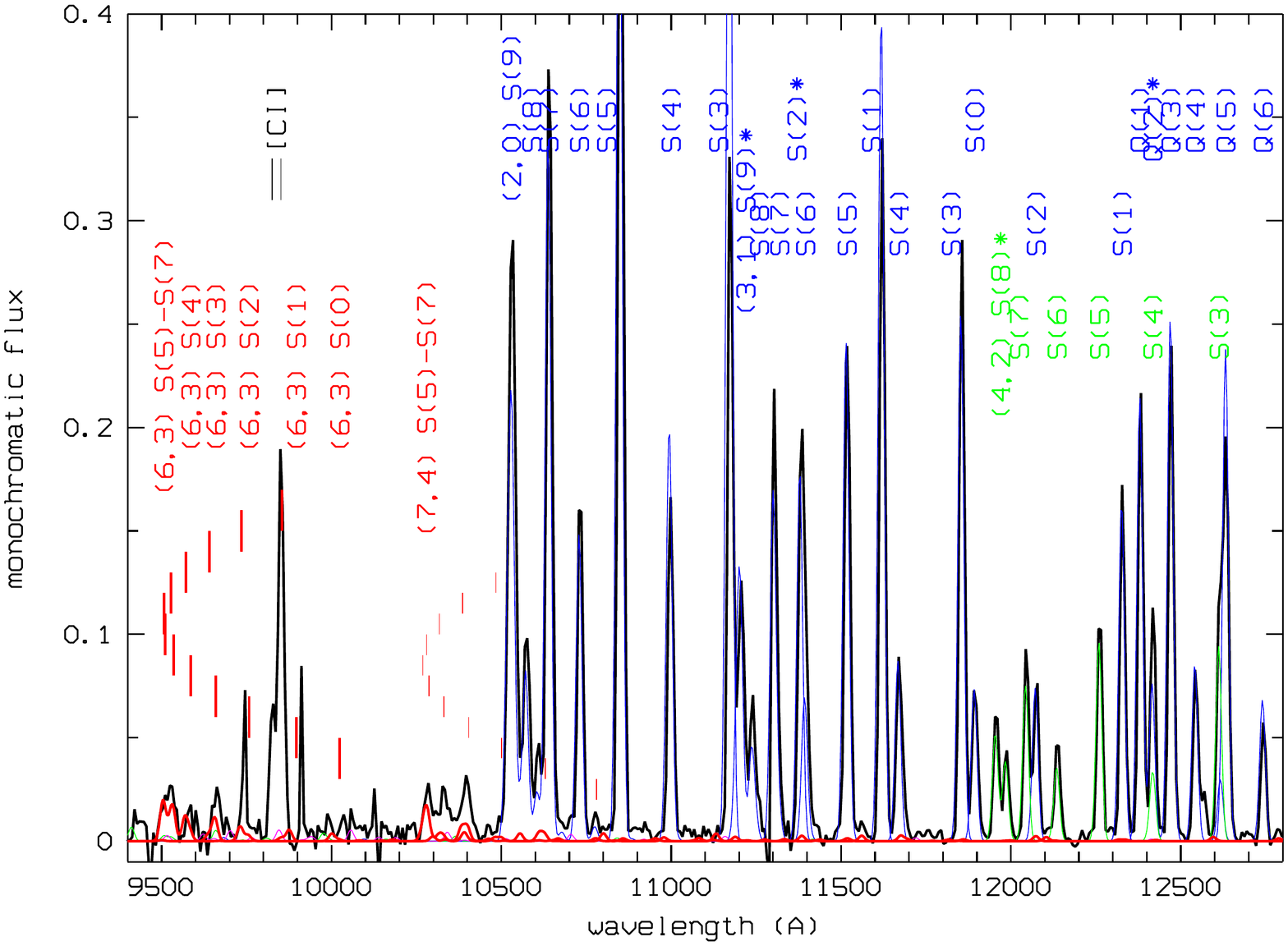}
      \caption{
J-band spectrum towards HH91A obtained with SOFI. Plotted are monochromatic fluxes in
units of $10^{-17}$W m$^{-2}\ \mu$m$^{-1}$ vs. wavelength in units of \ang.
The spectrum is dominated by strong
emission lines which arise from the (2,0), (3,1), and (4,2) S-branches. Faint emission 
from various lines in the (6,3) and (7,4) S-bands is detected. 
The position of the various lines
in the (6,3) S-branch between 9500--10\,000 \ang\ are indicated.
The position of several lines converging to the
band-head of the (7,4) S-branch between 10\,030--10\,050 \ang\ are indicated as well. 
The emission near at 9825\ang\ and 9851 \ang\ arises from the 
$^1$D$_2$--$^3$P$_1$ and $^1$D$_2$--$^3$P$_2$ transition of [CI]. The theoretical H$_2$
emission spectrum is reproduced in color, with emission from v'=2 and 3 in blue, 
from v'=4 in green, and from v'=6 and 7 in red.
              }
         \label{grb1}
   \end{figure*}

\begin{figure*}
   \centering
   \includegraphics[angle=-90,width=20cm]{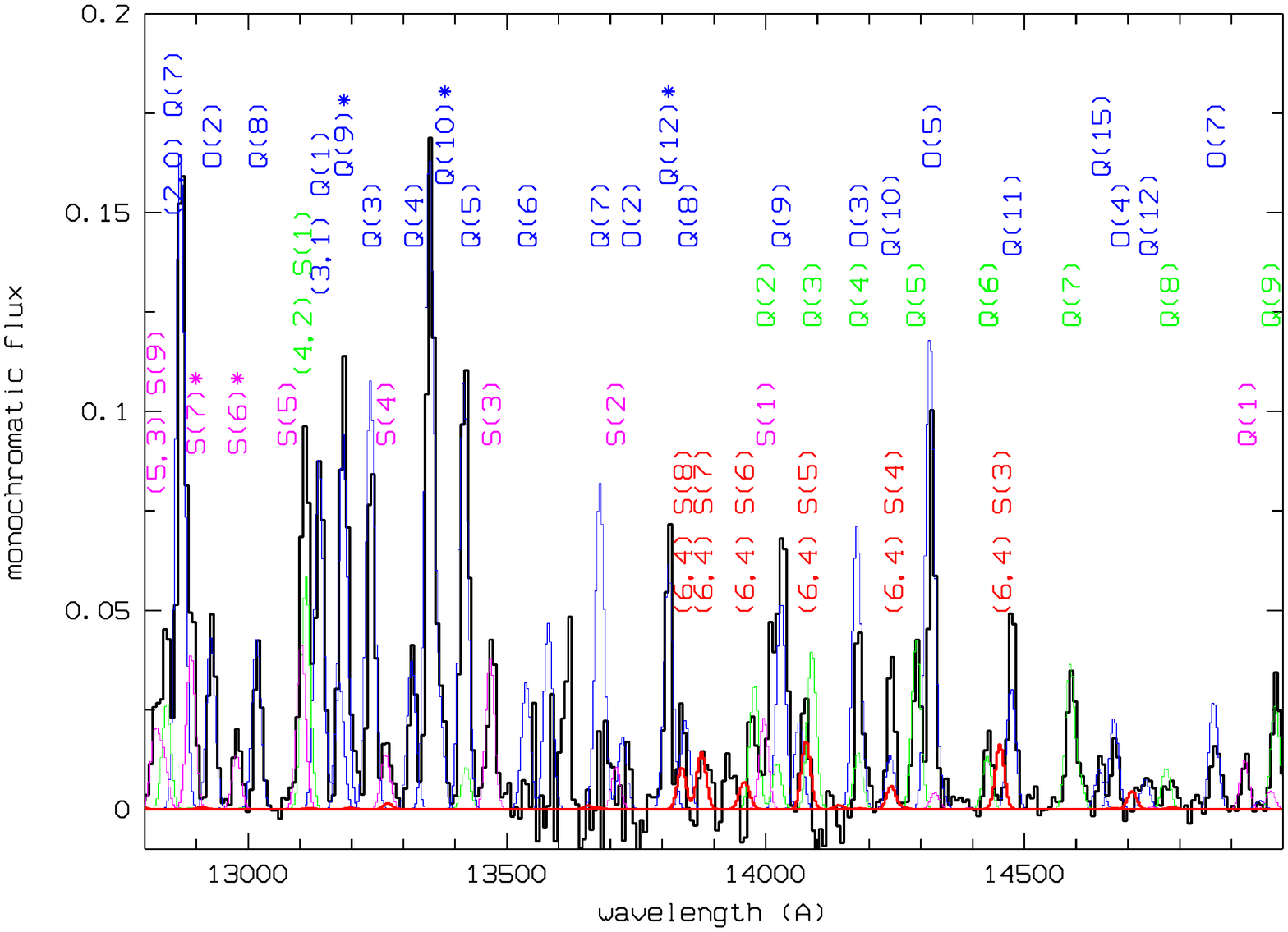}
      \caption{
J-band spectrum towards HH91A, with monochromatic fluxes in 
units of $10^{-17}$W m$^{-2}\ \mu$m$^{-1}$ vs. wavelength in units of \ang.
The strong emission lines which arise from the (2,0), (3,1), (4,2), and (5,3) bands
are identified. The expected position and strength of modeled emission from the
(6,4) S(3)--S(8) lines is indicated as well (cf. Sect.~\ref{discussion}). The
theoretical H$_2$ emission spectrum is reproduced in color, see Fig.~\ref{grb1}
for details.
              }
         \label{grb2}
   \end{figure*}

\begin{figure*}
   \centering
   \includegraphics[angle=-90,width=20cm]{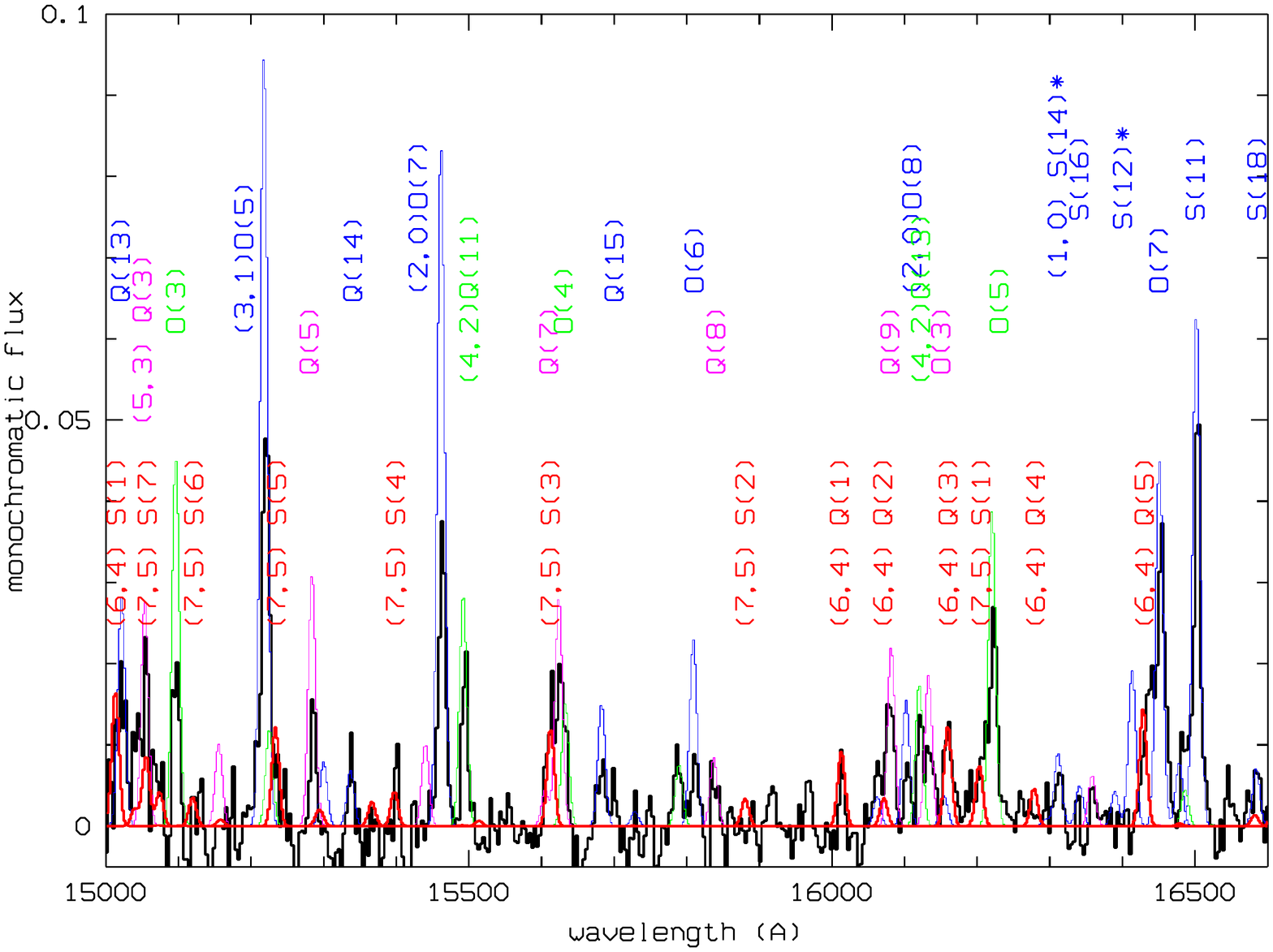}
      \caption{
H-band spectrum towards HH91A, with monochromatic fluxes in 
units of $10^{-17}$W m$^{-2}\ \mu$m$^{-1}$ vs. wavelength in units of \ang.
The emission lines reproduced in color correspond to a model
calculation which is presented in Sect.~\ref{discussion}. 
Emission from v'=1, 2, and 3 
in blue, v'=4 in green, v'=5 in magenta, and v'=6 and 7 in red.
It is noted that
emission from [FeII] $a^4\mathrm{D}_{7/2}-a^4\mathrm{F}_{9/2}$ near 16\,440 \ang\ is absent.
              }
         \label{Hhigh}
   \end{figure*}

\begin{figure*}
   \centering
   \includegraphics[angle=-90,width=20cm]{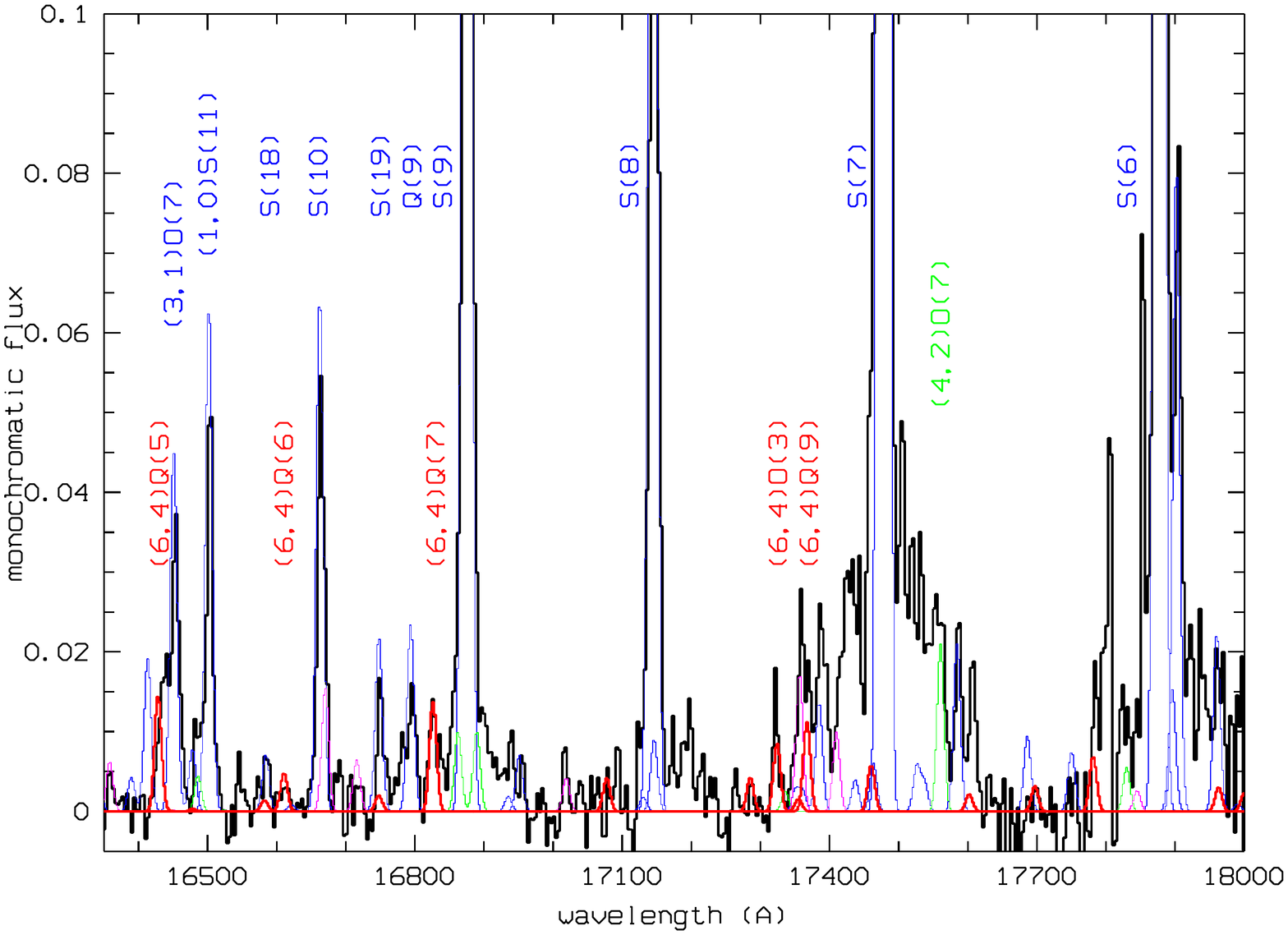}
      \caption{
H-band spectrum towards HH91A, see caption of Fig.~\ref{Hhigh} for details.
              }
         \label{H}
   \end{figure*}

\begin{figure*}
   \centering
   \includegraphics[angle=-90,width=20cm]{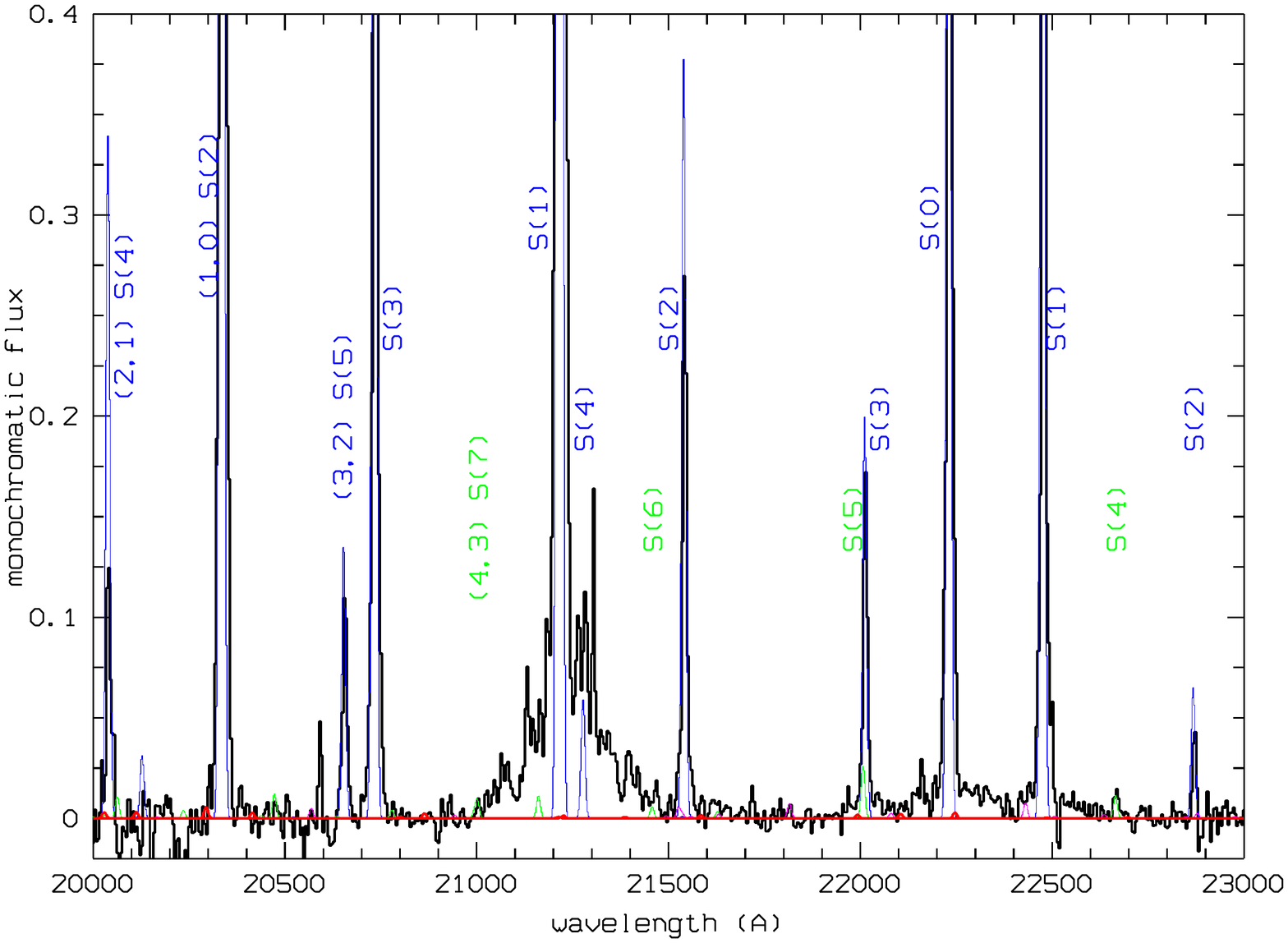}
      \caption{
Ks-band spectrum towards HH91A obtained with SOFI, see caption of 
Fig.~\ref{Hhigh} for details. The wings in the (1,0) S(1) line arise
from an instrumental defect of SOFI and have no astrophysical significance.
              }
         \label{ks}
   \end{figure*}

The near-infrared spectra obtained with SOFI are reproduced in Figs.~\ref{grb1}--\ref{ks}.
The J-band spectra shown in Figs.~\ref{grb1},\ref{grb2} are dominated by 
emission from the (2,0) S-branch (band-head near 1.055 $\mu$m), the (3,1) S-branch (band-head
near 1.118 $\mu$m), the (4,2) S-branch (band-head near 1.185 $\mu$m), and the (5,3) S-branch
(band-head near 1.282 $\mu$m). Strong emission lines from the (2,0), (3,1), and (4,2) Q-branch
is detected as well. In addition, various emission lines from the (6,3) S-branch band are
detected long-ward of its band-head near 9506 \ang, and from the (7,4) S-branch (band-head
at 10028 \ang), are detected (cf. Fig.~\ref{grb1}). The inferred population densities
in the ro-vibrational levels of v'=6 imply that emission in the (6,4) band occur at
flux levels above the noise of the spectra presented here. The (6,4) Q(1)--Q(9) lines
are clearly detected (see below). 
The expected emission lines in (6,4) S-branch, up to the band-head marked by the 
(6,4) S(8) line, near 13840 \ang, is reproduced in Fig.~\ref{grb2} by the red line. 
The (6,4) S-branch is located in a region of poor atmospheric transmission between 
1.35--1.5 $\mu$m, where the fluxes of the measured emission lines are highly uncertain.
The modeled emission in the (6,4) S-branch is consistent with the observations.
The emission feature near 9825--9851 \ang\ corresponds to emission from atomic carbon
(cf. Sect.~\ref{discussion}). Emission from [FeII], which is generally observed in
HH-objects, is absent. 

The H-band spectra towards HH91A are shown in Figs.~\ref{Hhigh},\ref{H}. The H-band spectrum
is dominated by strong emission from the (1,0) S-branch (band-head marked by 
(1,0) S(14) near 16296 \ang) and relatively strong emission from the (3,1) O(5)--O(7),
(4,2) Q(11)--Q(13). In addition, emission from (6,4) Q(1)--Q(9) is detected.
The bold red line reproduced in Figs.~\ref{Hhigh},\ref{H}
correspond to the theoretical emission from a model presented below.


Finally, the Ks-band spectrum is shown in Fig.~\ref{ks}. The emission is dominated by the
very strong (1,0) S(0)--S(2) lines and the (2,1) S(1)--S(4) lines. Emission from (3,2) 
S(2)--S(5) and from (4,3) S(4) is detected as well. {The strong H$_2$ lines such as
the (1,0) S(7) and the (1,0) S(1) line show pronounced line wings. Those wings
have no astrophysical significance and arise from an instrumental defect of SOFI, 
evidenced by the fact that these lines do not show wings in the spectra taken 
previously with IRSPEC (Gredel et al. 1992).} 

{The spectra shown in Figs.~\ref{red1}--\ref{ks} contain some 200 emission lines of
molecular hydrogen. The inferred fluxes $F$ of the various lines 
are given in column 3 of Tables~\ref{tableone} and \ref{tabletwo}, with flux 
uncertainties in parenthesis.
The optical spectra have limiting line fluxes of about 
$0.5 \times 10^{-19}$W m$^{-2}$, as
judged from noise in the flux-scaled spectra (see above). Limiting fluxes are
about $10^{-19}$ W m$^{-2}$ in the spectra taken with grism GB, $0.5 \times 10^{-19}$
W m$^{-2}$ in the H-band and $10^{-19}$ W m$^{-2}$ in the Ks-band
taken with grism HR. For the stronger lines, flux uncertainties introduced by the
calibration of the atmospheric transmission are estimated to be of the order of
10--20\% of the total line flux. 
Fluxes derived from emission lines which occur in spectral
regions which are dominated by  narrow, telluric absorption lines, such as the 
13\,500--15\,000\ang\ region, are uncertain by larger amounts. 
Columns 1, 2, and 4 contain the line identification, the vacuum wavelength $\lambda$, 
and the inferred column density $N(\mathrm{v'J'})$ 
of the corresponding upper ro-vibrational level v'J', respectively. 
Numbers in parenthesis in column 4 of Tables~\ref{tableone} and \ref{tabletwo} 
are uncertainties in column densities.}

Because of the relatively low spectral resolution provided by SOFI, many
of the features detected in the J-, H-, and Ks-bands are blends of two
or more emission lines of H$_2$. In cases of line blends, no effort
was made to de-convolve the lines or to assign fractional flux values to the 
individual components. Rather, the full line flux was assigned to each one 
of the possible ro-vibrational lines which occur at the given
wavelength. Examples are  
the (4,2) S(9) + S(10) blend near 1.196 $\mu$m,
the (2,0) Q(2) + (4,2) S(4) blend near 1.242 $\mu$m,
the (4,2) S(2) + (5,3)S(10) blend near 1.284 $\mu$m,
or the (2,0) Q(7) + (5,3)S(7) blend near 1.288 $\mu$m. 
The entries in Table~\ref{tabletwo} are thus to be read with care - in cases when line
blends occur, the listed H$_2$ column densities are too large by factors of a few. 
Unresolved line blends are explicitly identified in Figs.~\ref{grb1}--\ref{ks} 
and in Tables~1 and by an asterisk.

\begin{figure}
   \centering
   \includegraphics[angle=0,width=10cm]{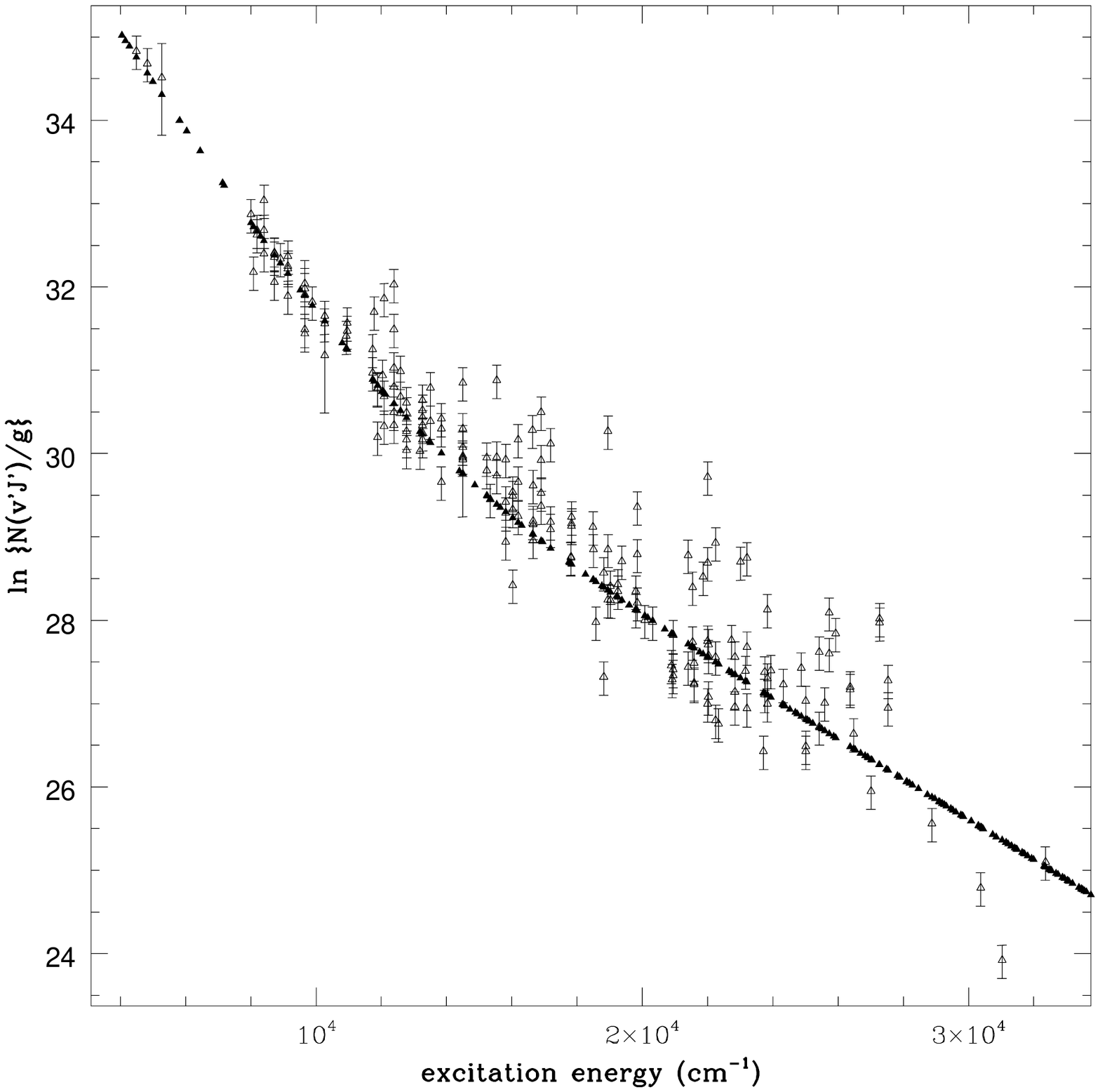}
      \caption{
H$_2$ excitation diagram with values of $ln N$(v'J')/g plotted versus excitation energy
$E$(v'J'). Open triangles represent data points inferred from the
column densities $N$(v'J') listed in Tables~1 and 2.
Filled triangles are obtained from a model calculation where the emission arises 
from a two-component gas model, where the bulk of the material 
has a total H$_2$ column density of $1.24 \times 10^{18}$ cm$^{-2}$ and is at a temperature 
of 2750~K, and where a fraction of $10^{16}$ cm$^{-2}$ of H$_2$ is at a temperature of 6000~K. 
              }
         \label{exc}
   \end{figure}

The H$_2$ column densities listed in Tables~1 and 2 are nevertheless 
used to construct the H$_2$ excitation diagram with values of
$ln(N(\mathrm{v'J'})/g)$ plotted versus excitation energy $E(\mathrm{v'J'})$ (see eg. Gredel
\cite{gredel06}
for details). The diagram is reproduced in Fig.~\ref{exc}. The occurrence of
line blends introduces some scatter to the
excitation diagram. The scatter has no physical origin nor is it introduced by 
non-thermal excitation scenarios. This statement is justified in detail in 
Sect.~\ref{discussion} below.  We proceed in the 
following iterative way to derive the ro-vibrational excitation temperature 
of the v'J' levels. The population
densities in the H$_2$ levels up to an excitation energy of say $10^4$ cm$^{-1}$
is consistent with an excitation temperature of 2750~K, which is the temperature derived
by Gredel et al. \cite{gredel92} from their IRSPEC spectra which were
obtained at a higher spectral resolution than the spectra presented here.
The population densities among the ro-vibrational levels 
above excitation energies of $10^4$ cm$^{-1}$ deviate from the population densities
expected for a thermalized distribution at 2750~K. 
The deviation causes a {\it curvature} in the excitation diagram 
and indicates that a fraction of 1\% of the 
gas is at the very high temperature of 6000~K. This statement assumes that 
all the levels up to excitation energies
of about 30\,000 cm$^{-1}$, or some 40\,000~K, are thermalized. Higher gas-kinetic
temperatures are in principle possible if it is assumed that the
levels are sub-thermally excited. 
The curvature in the H$_2$ excitation diagram is not very pronounced and is only 
established through the observation of high-excitation emission lines 
with excitation energies above 15\,000 cm$^{-1}$. This is
the reason why the curvature went unnoticed in the earlier work of
Gredel et al. (\cite{gredel92}). The relatively low degree of temperature
stratification in HH91A, combined with the absence of emission from [FeII],
supports the general finding of Caratti o Garatti et al. (2006) who concluded
that a significant temperature stratification in the H$_2$ emitting gas is in general
observed in HH-objects which show [FeII] emission as well. 

In order to judge whether the above conclusions are fully consistent with the
observed optical and near-infrared spectra, we have modeled the spectrum expected
from a two-component gas mixture with is at a temperature of 2750~K and where a 
fraction of 1\% of the gas is at a temperature of 6000~K. We have used the models
of Gredel \& Dalgarno (1995) to calculate the theoretical H$_2$ emission spectrum. From 
the entry rates into the ro-vibrational levels v'J' of the electronic ground state of 
H$_2$, a spectrum (Voigt profiles) is calculated as a function of parameters such
as the total H$_2$ column density, the reddening $E_\mathrm{B-V}$, the desired
spectral resolution R = $\lambda/\Delta \lambda$, etc. We ignore
reddening towards HH91A (cf. Gredel et al. 1992) and calculate the expected 
emission spectra for the various spectral resolutions in the PMAS and SOFI spectra. 
Apart from the relative flux calibration of the PMAS spectra as discussed
above, the only scaling that we involve is introduced by a forced match of the 
predicted and calculated flux in the (1,0) S(1) line. 
For a gas mixture of warm molecular gas at 2750K plus a fraction of 1\% at 6000~K,
the model calculation produces a total H$_2$ flux of
$F_\mathrm{tot}$(H$_2) = \Sigma_\mathrm{v'J'v''J''}
F\mathrm{(v'J'v''J'')} = 19.8 \times F\mathrm{(1301)}$,
where $F\mathrm{(1301)}$ is the flux in the (1,0) S(1)
line and where the summation is carried out over all possible emission lines of H$_2$.
The total H$_2$ population density, for the two-component gas mixture adopted here,
is $N_\mathrm{tot}(\mathrm{H}_2) = \Sigma_\mathrm{v'J'} N\mathrm{(v'J')}
= 45.7 \times N\mathrm{(1,3)}$, or 
$N_\mathrm{tot}(\mathrm{H}_2) = 1.26 \times 10^{18}$ cm$^{-2}$.
The scaling factors (19.8 and 45.7 in the present case) are strongly 
dependent on the temperatures and column density ratios (cf. Gredel \cite{gredel94}).

The model calculations are reproduced by the colored lines in Figs.~\ref{red1}--\ref{ks}.
In order to illustrate the contributions from the various vibrational levels of
H$_2$, emission which arises from vibrational levels v'$\le$3 is color coded in blue, 
emission from v'=4 in green, emission from v'=5 in magenta, and emission from v'$\ge$6
in red. The agreement of the modeled spectrum with the observations is excellent. 
In general, the line fluxes in the 200 or so observed emission lines of H$_2$ are
reproduced within 20\%. Relatively large
deviations (factor of 2) between the model spectrum and the observations occur for 
the (3,0) S(1) line near 8150\ang, for (4,1) S(1) and (3,0) Q(5) near 8670 \ang, 
and for (3,1) Q(7) near 1.37 $\mu$m, (3,1) O(5) near 1.523 $\mu$m, and (2,0) O(7) near
1.545 $\mu$m.  The model also fails to reproduce the observed emission features 
near 1.03$\mu$m but it does reproduce the (7,5) S-branch
in the H-band, which contains lines which arise from the same upper ro-vibrational
levels than the lines near the (7,4) S-branch band-head near 1.03 $\mu$m. This
may indicate that emission other than from the (7,4) S-branch occurs near 1.03$\mu$m.
The model spectrum demonstrates that fluxes of a few  $10^{-20}$ W m$^{-2}$
of individual ro-vibrational lines in the (6,3), (6,4), (7,4), and (8,4) bands are
expected from a thermally excited gas towards HH91A.
Among the various high-excitation lines, the model reproduces perfectly 
well the band head of the (6,3) S-branch near 9500 \ang, the (6,4) Q(1)--Q(9)
lines in the H-band. 
Given that some of the discrepant lines occur in relatively poor atmospheric
windows, and that the rest of the 200 or so observed lines are very well reproduced,
and that fluxes among lines that arise from ro-vibrational levels that span
excitation energies from 4000--30\,000 cm$^{-1}$ are accurately modeled, 
we ignore the discrepancies and conclude that the observed H$_2$ emission
arises from thermal gas at 2750~K which contains a fraction of 1\% at a temperature
of 6000~K.

{Tables ~3 and 4 contains a full listing of our model results, and gives 
expected H$_2$ emission lines which have integrated line fluxes
above $F_\mathrm{tot} = 10^{-19}$ W cm$^{-2}$ (Table~3) and fluxes ranging
between $0.5 - 1 \times 10^{-19}$ Wm$^{-2}$ (Table~4).
The predicted thermal fluxes towards HH91A
from the two gas components (bulk at 2750~K and 1\% at 6000~K)
are listed separately in columns 4 and 5, respectively. The tables contain the line 
identification, the wavelength in $\mu$m, and the energy of the upper ro-vibrational 
level in cm$^{-1}$, in columns 1--3, respectively. Total line fluxes are given
in column 6. Figures~\ref{resred1}--\ref{resKs} contain the residuals between
the observed and the modeled H$_2$ line fluxes. As discussed earlier in Sect.~\ref{results},
it can be seen that the overall agreement  between the observed and the modeled spectra 
is excellent.}

\section{Discussion}
\label{discussion}

Optical and near-infrared emission lines from molecular hydrogen
which arise from very high-excitation ro-vibrational levels
in the electronic ground state of H$_2$ are generally seen in sources where
electronic states of H$_2$ are pumped in strong ultraviolet radiation fields, such as in 
NGC~2023 (McCartney et al. \cite{McCartney}). The absorption of ultraviolet radiation
in the Lyman and Werner bands of H$_2$ and the subsequent decay of the 
excited electronic states via dipole radiation populates the 
ro-vibrational levels v'J' of the electronic X$^1\Sigma_g^+$ ground state
of H$_2$. The excited v'J' levels cascade 
to lower ro-vibrational levels v''J'' via electric quadrupole (E2) radiation and 
give rise to optical and near-infrared emission of H$_2$. 
In regions with strong X-ray radiation fields, electronic states of H$_2$ may
also be collisionally excited by energetic secondary electrons produced by
X-ray ionizations (Gredel \& Dalgarno 1995). X-rays have been detected in
very fast shocks in HH-objects (HH2A, Pravdo et al. 2001; HH 154, Bally, Feigelson, \&  
Reipurth 2003). 
Pumping by Ly$\alpha$ photons of H$_2$ is possible in
a warm gas which contains a fraction of H$_2$ in the v'=2,J'=5 level (Schwartz et al.
\cite{schwartz}). The excitation of electronic states by UV or Ly$\alpha$ photons
follows dipole selection rules while the collisional excitation by secondary electrons
does not. Non-thermal excitation of the ro-vibrational levels may also occur in a gas
where H$_2$ reforms after the passage of a strong, dissociative shock 
(LeBourlot et al. (1995); Casu \& Cecchi-Pestellini \cite{casu}; 
Tin\'e et al. \cite{tine03}). In such models,
uncertainties about whether the H$_2$ formation energy is equipartitioned among 
the ro-vibrational levels, the kinetic energy of the molecule, and the internal energy
of the grain lattice, translate to significant differences in the modeled H$_2$
spectra, and renders a comparison with the observations difficult.

The shocked gas in molecular outflows from protostars may be affected by
the non-thermal excitation scenarios described above. Fast, dissociative shocks 
produce a radiative precursor which contains a strong ultraviolet radiation field. 
Embedded TTau stars in the star forming regions may contribute a significant X-ray
radiation field to the environment (Guedel et al. 2007, and references therein).
Non-thermal excitation scenarios introduce a pronounced
dependence of the rotational and vibrational excitation temperatures of the
ro-vibrational levels in the electronic ground state of H$_2$. Such is evidenced
by strong deviations from the smooth Boltzmann distribution which characterizes 
thermal gas. 

None of these effects dominate the excitation of H$_2$ in HH91A. 
What is immediately clear from the H$_2$ excitation diagram shown in Fig.~\ref{exc},
but more convincingly from the excellent agreement of the 
model H$_2$ emission spectra which arises from thermally excited gas and the observations, 
is that all ro-vibrational levels up to excitation energies of 40\,000~K are in 
LTE.  The fluxes in the high-excitation emission lines in the observed (6,4), (7,4), and (8,4)
bands are in excellent agreement with the expected strengths from the two-component
thermal gas described above. The presence of these lines 
does not require to employ non-thermal excitation scenarios to explain
the observed H$_2$ emission towards HH91A. 

At kinetic temperatures of a few 1000~K, the rate coefficients for collisional
excitation of H$_2$ by hydrogen atoms are of the order of $10^{-12}$ cm$^{-3}$ s$^{-1}$. 
In order to estimate the extent at which the non-thermal excitation scenarios discussed
above contribute to the H$_2$ excitation, we use the models of Gredel \& Dalgarno
(1995) to calculate the entry rates into the ro-vibrational levels of the ground
state from X-ray and UV-fluorescence. X-ray ionization rates need to be 
significantly larger than
$\zeta = 10^{-15}$ s$^{-1}$ for collisional impact excitations of electronic H$_2$ 
states by fast secondary electrons to result in entry rates which exceed those from 
thermal excitations in a gas of a few 1000~K temperature. 
The generally adopted value of the cosmic-ray ionization rate in dense gas is
$\zeta = 10^{-17}$ s$^{-1}$. The upper limit to the ionization fraction from X-rays
is about $x_\mathrm{e} \le 10^{-4}$. It can thus be ruled out that X-rays
contribute significantly to the H$_2$ emission observed towards HH91A. 
Similarly, a strong ultraviolet radiation field which exceeds the strength of
the ambient interstellar radiation field by factors of several 100 is 
required for UV fluorescence to compete with the thermal population of H$_2$ levels
in the ground state. The presence of a fast, dissociative shock with a strong
UV precursor should thus be ruled out as well towards HH91A.

The fact the H$_2$ levels up to excitation energies of 30\,000 cm$^{-1}$ are 
thermalized requires very large densities in the compressed, post-shock gas. 
The critical densities which are required to populate the ro-vibrational levels are 
equal to the Einstein A-values divided by the collisional de-excitation
rate coefficients, $ n_\mathrm{crit}$(v'J')=$A$(v'J'v''J'')/$<\sigma v>$. 
The critical densities exceed values of $n_\mathrm{crit} > 10^7$ cm$^{-3}$ 
for levels with excitation energies above 30\,000 cm$^{-1}$. 
A more careful inspection of the H$_2$ excitation diagram shown in Fig.~\ref{exc}
shows that the population density among some of the very high-excitation levels, 
with excitation energies above 30\,000~K, may show signs of sub-thermal excitation. 
In particular, the population density inferred from the optical (7,3) S(11) line
at 7978 \ang, which has an excitation energy of 30\,368 cm$^{-1}$, has a measured
flux which is about a factor of three lower than what is expected from the 
our two-component thermal model. This may indicate the onset of 
subthermal excitation for the very high levels such as the $v'=7,J'=13$
level from which the (7,3)S(11) line arises.

From the optical observations of [SII] 6717\ang\ and 6731\ang\ lines, Gredel et al. (1995) 
inferred very low electron densities of the order of $n_e \approx 300$
cm$^{-3}$ towards HH91A. This finding is consistent with the upper limit
of the ionization fraction of $10^{-4}$ and the critical densities
derived above. The absence of emission from 
[FeII] (e.g.  the $a^4\mathrm{D}_{7/2} - a^4\mathrm{F}_{9/2}$ near 1.644$\mu$m) supports
the idea that H$_2$ is excited by a relatively slow, non-dissociative shock. 
Emission from ionized atomic species such as [FeII] is often observed in
Herbig-Haro objects (eg. Nisini et al. \cite{nisini}), yet the strength 
of the atomic lines is 
not well reproduced by shock-models which explain the H$_2$ emission. This 
suggests that [FeII] arises from faster, dissociative shocks and in
regions which are distinct from H$_2$-emitting regions. 
Weak emission from [CI]
is seen near 8727 \ang, with a flux of $F_{8727} \approx 0.8 \times 10^{-19}$ W m$^{-2}$,
and near 9825 \ang\ and 9851 \ang\ of $F_{9825} = 10.8 \times 10^{-19}$ W m$^{-2}$ and
$F_{9851} = 37 \times 10^{-19}$ W m$^{-2}$. The observed [CI]8727/(9825+9851) line ratio 
of 0.02 is well reproduced by slow, non-dissociative shocks. We thus conclude that
the emission seen towards HH91A is produced in a slow J-type shock. This picture
is in agreement with the expectations that evolved outflows favor the formation
of J-shocks (Caratti o Garatti 2006). 
It has been pointed out by Flower et al. (\cite{flower}) that the discrimination
between C-type and J-type shocks based on H$_2$ excitation diagrams is far from
straightforward. A J-shock is preferred here because C-type shocks fail, in general,
to produce the high degree of thermalization that is observed here. {This conclusion
is in agreement with earlier results by Smith (1994) whose analysis is based 
on fewer observed H$_2$ lines.} 

\begin{figure*}
   \centering
   \includegraphics[angle=-90,width=20cm]{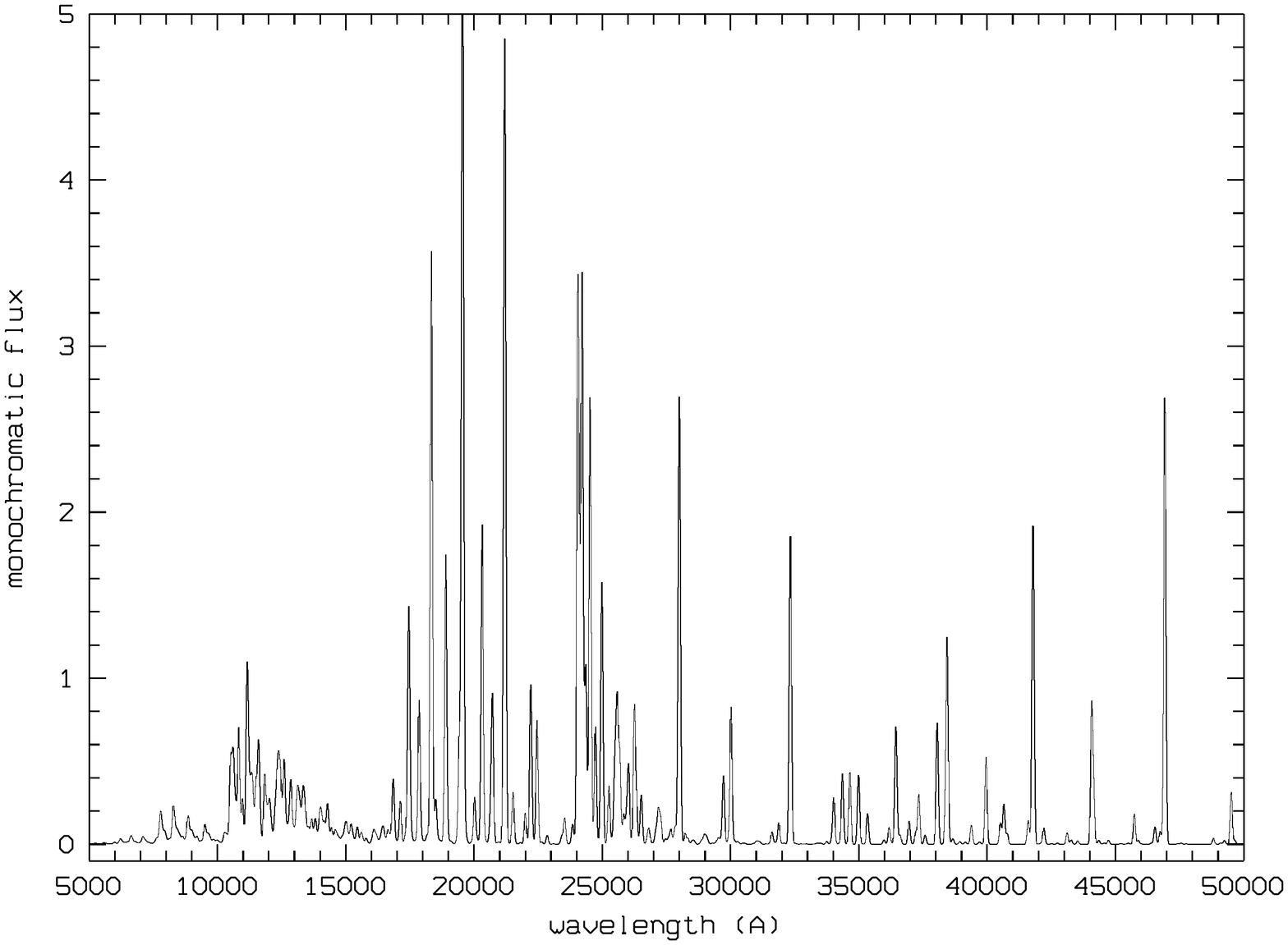}
   \caption{
Modeled H$_2$ emission spectrum which results from a total column of N(H$_2) = 
10^{18}$ cm$^{-2}$ of hot gas at a temperature of 2750~K, which contains a fraction 
of 1\% of hot gas at 6000~K. Fluxes are in units of $10^{-19}$W m$^{-2}$ and 
wavelengths are in \ang. Emission lines are Voigt line profiles at a FWHM of 200 \ang.
               }
              \label{full}
    \end{figure*}

\begin{figure*}
   \centering
   \includegraphics[angle=-90,width=20cm]{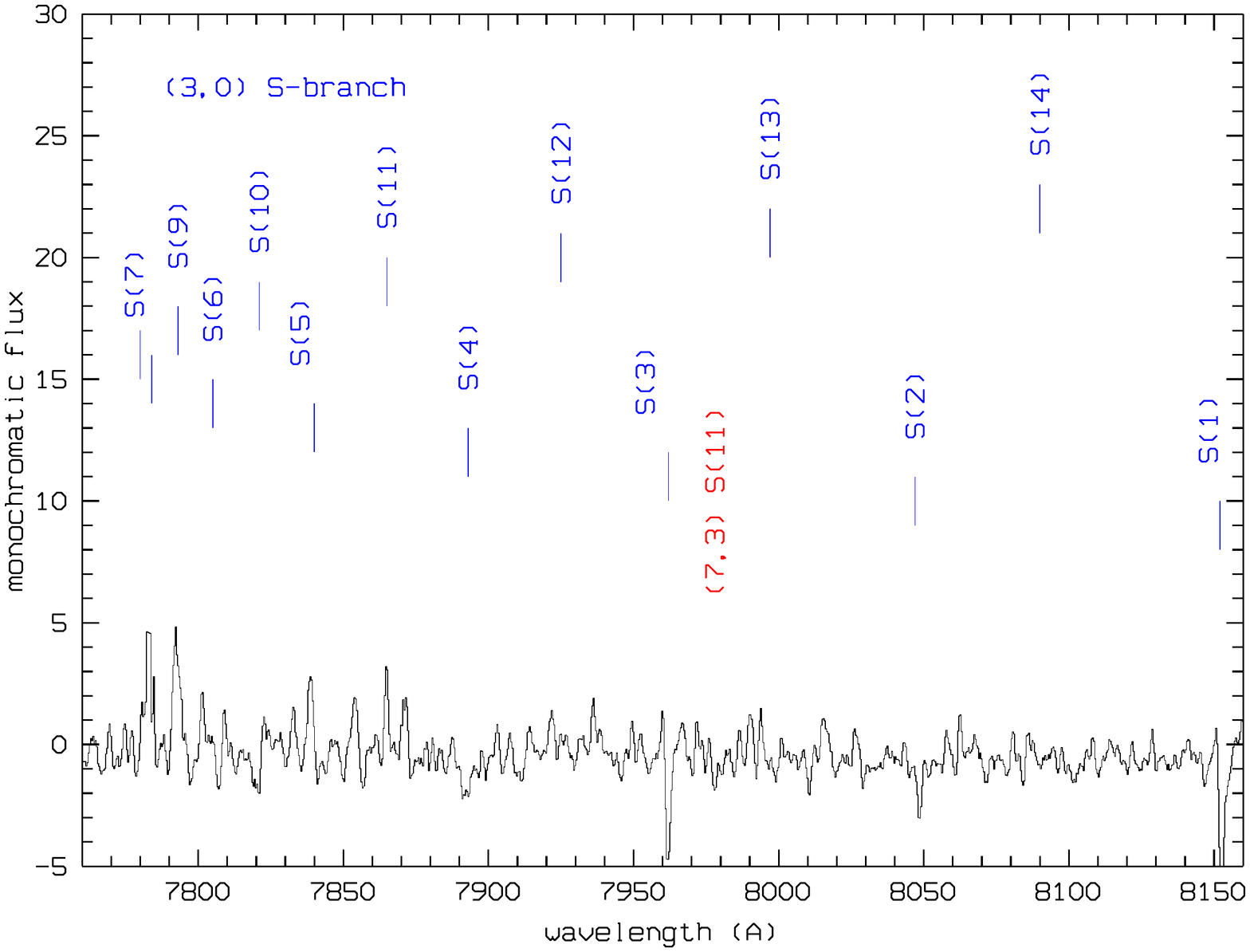}
   \caption{
Residuals between observed and modeled H$_2$ spectra as shown in Fig.~1, 
covering the range of 7760--8160 \ang. The expected position 
of the (3,0) S(1)--S(14) lines are indicated.
               }
              \label{resred1}
    \end{figure*}

\begin{figure*}
   \centering
   \includegraphics[angle=-90,width=20cm]{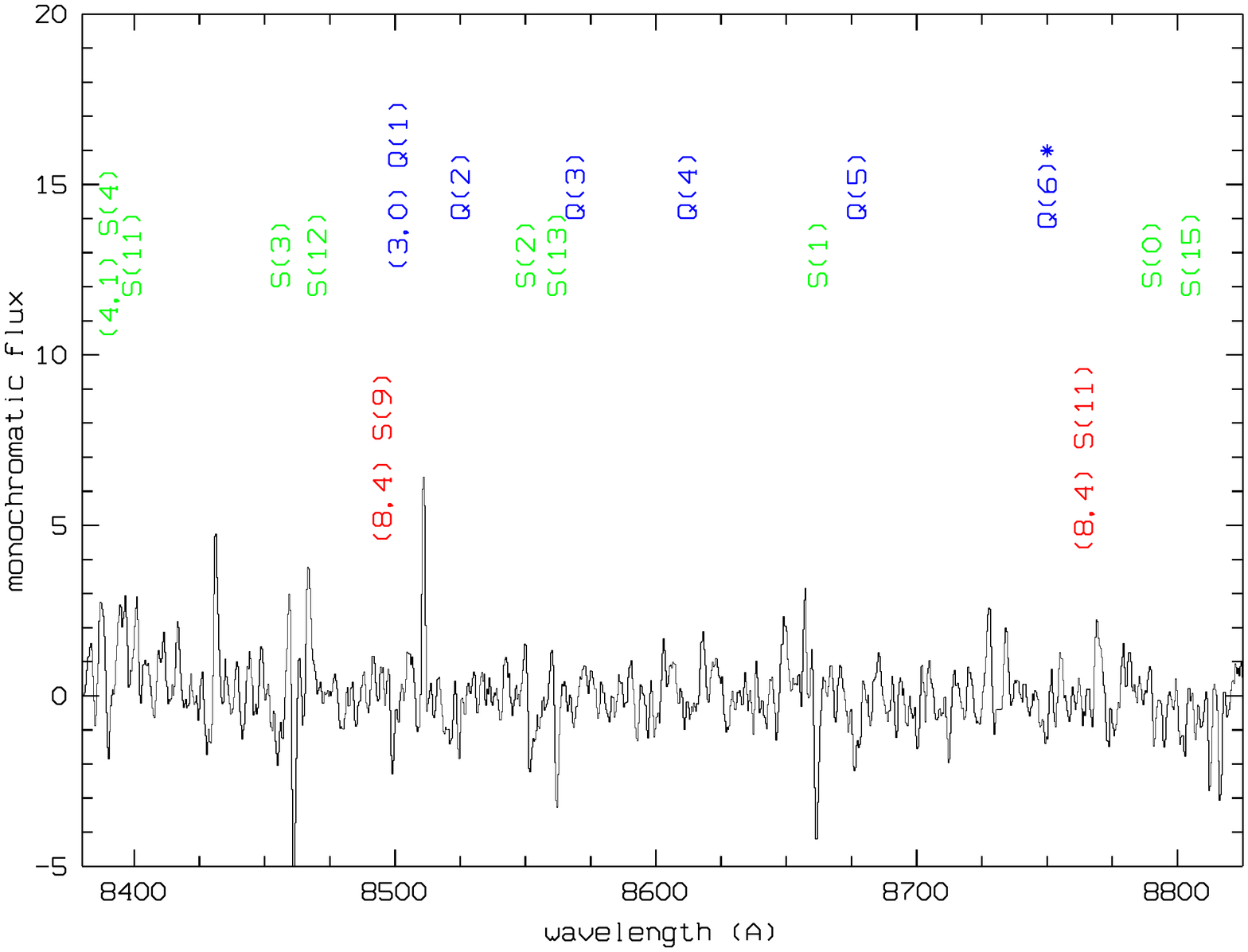}
   \caption{
Residuals between observed and modeled H$_2$ spectra as shown in Fig.~2,
covering the range of 8390--8820 \ang. The expected position
of various emission lines in the (3,0) and (4,1) bands are indicated.
The feature near 8510 \ang\ is a spike and does not
correspond to an H$_2$ emission line.
               }
              \label{resred2}
    \end{figure*}

\begin{figure*}
   \centering
   \includegraphics[angle=-90,width=20cm]{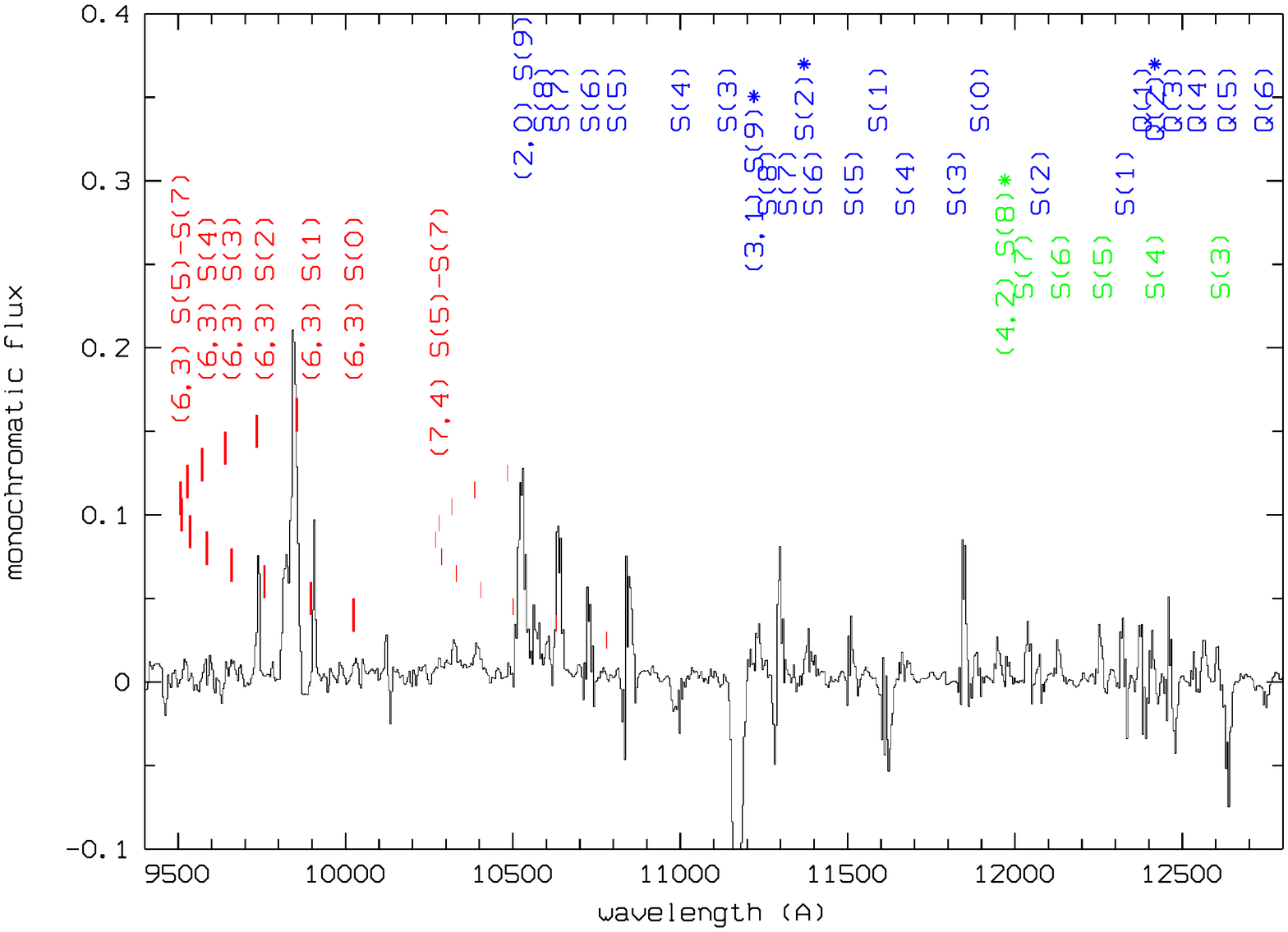}
   \caption{
Residuals between observed and modeled H$_2$ spectra as shown in Fig.~3, 
covering the range of 9400--12800 \ang. The modeled flux in the (3,1) S(9),
S(10), S(11) blend near 11\,200 \ang\ is too high by about a factor of 2.
               }
              \label{resgrb1}
    \end{figure*}

\begin{figure*}
   \centering
   \includegraphics[angle=-90,width=20cm]{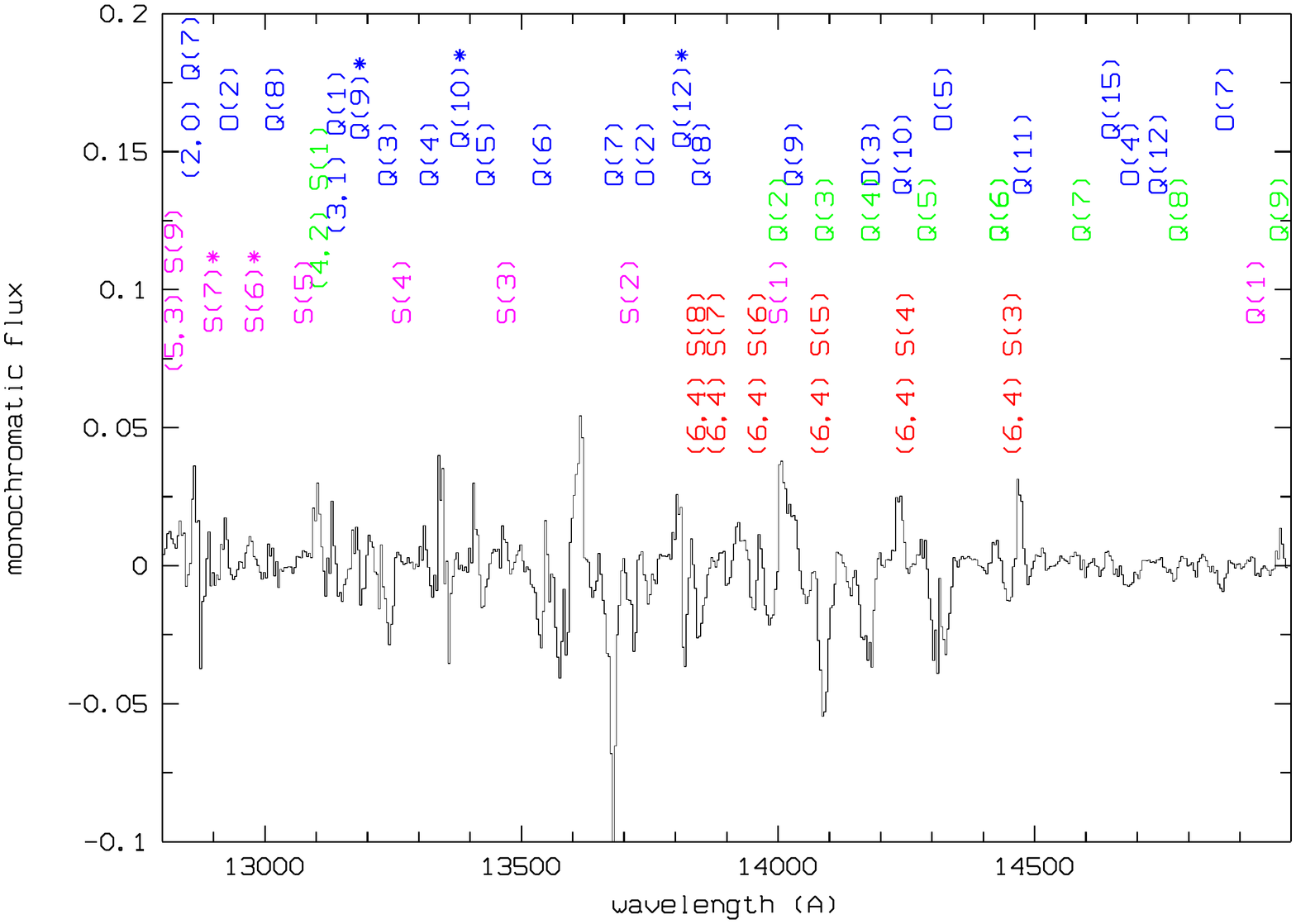}
   \caption{
Residuals between observed and modeled H$_2$ spectra as shown in Fig.~4,
covering the range of 12\,800--15\,000 \ang. The modeled flux in the (3,1) Q(7) line
is too high by about a factor of 2. The spectral region between 13\,500--14\,500 \ang\ 
is characterised by poor atmospheric transmission.
               }
              \label{resgrb2}
    \end{figure*}

\begin{figure*}
   \centering
   \includegraphics[angle=-90,width=20cm]{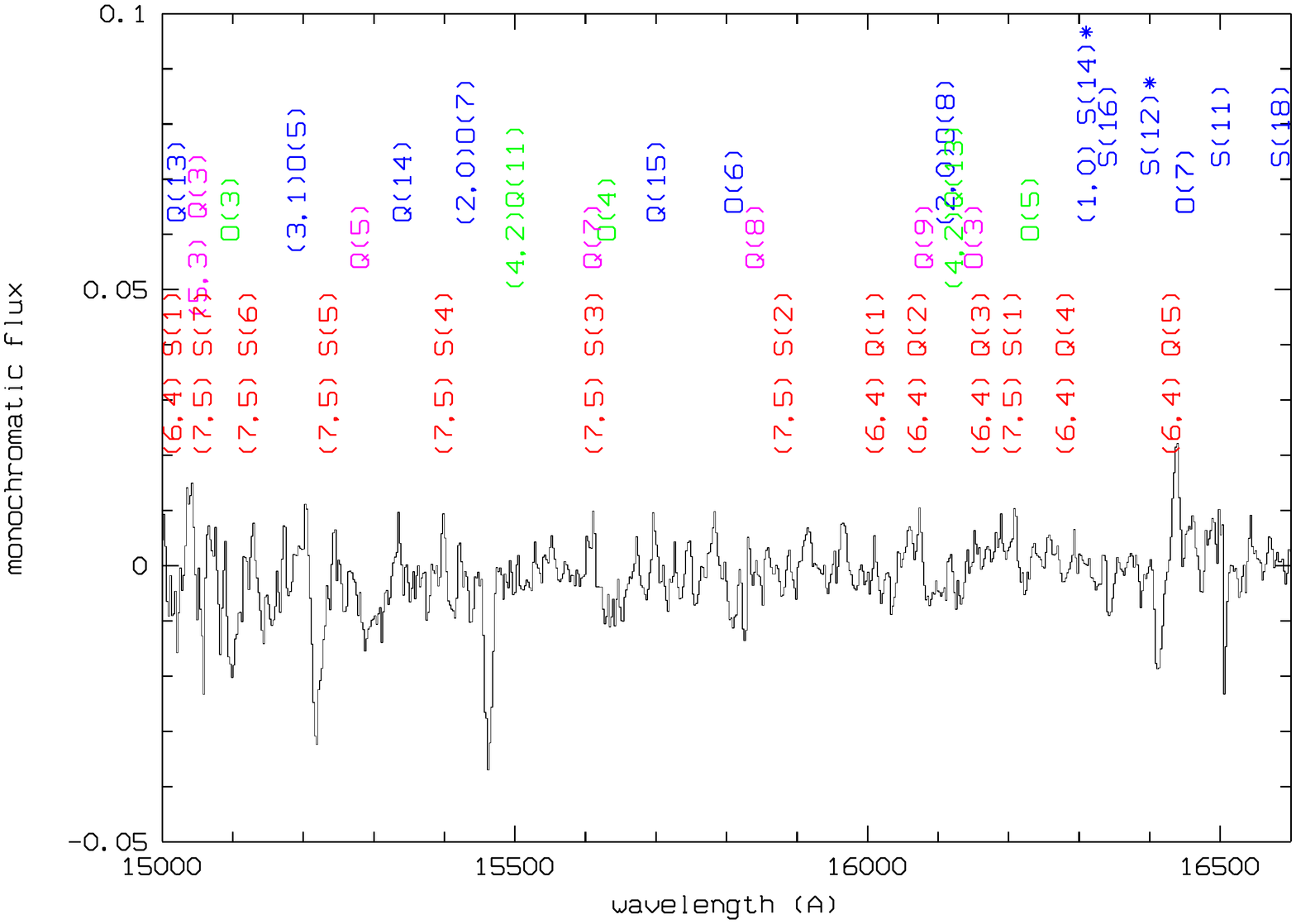}
    \caption{
Residuals between observed and modeled H$_2$ spectra as shown in Fig.~5, 
covering the range of 15\,000--16\,600 \ang. The flux in the (3,1) O(5) and (2,0) O(7)
lines is too strong by about a factor of 2.
               }
              \label{resH_high}
    \end{figure*}

\begin{figure*}
   \centering
   \includegraphics[angle=-90,width=20cm]{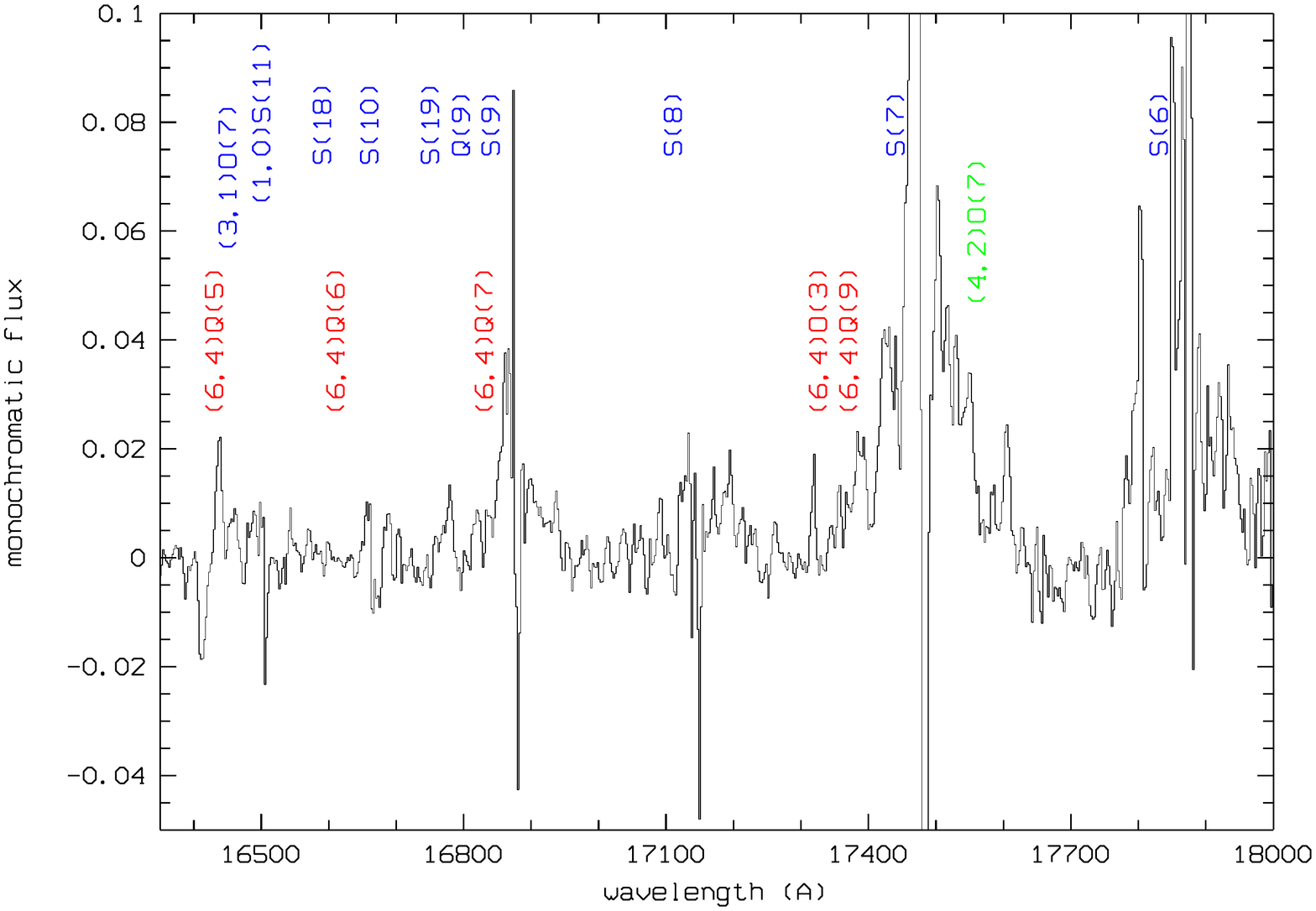}
   \caption{
Residuals between observed and modeled H$_2$ spectra as shown in Fig.~6,
covering the range of 16\,350--18\,000 \ang. The line wings in the (1,0) S(7) 
and S(8) lines show line wings arise from an instrumental defect of SOFI
and have no astrophysical significance.
               }
              \label{resH}
    \end{figure*}

\begin{figure*}
   \centering
   \includegraphics[angle=-90,width=20cm]{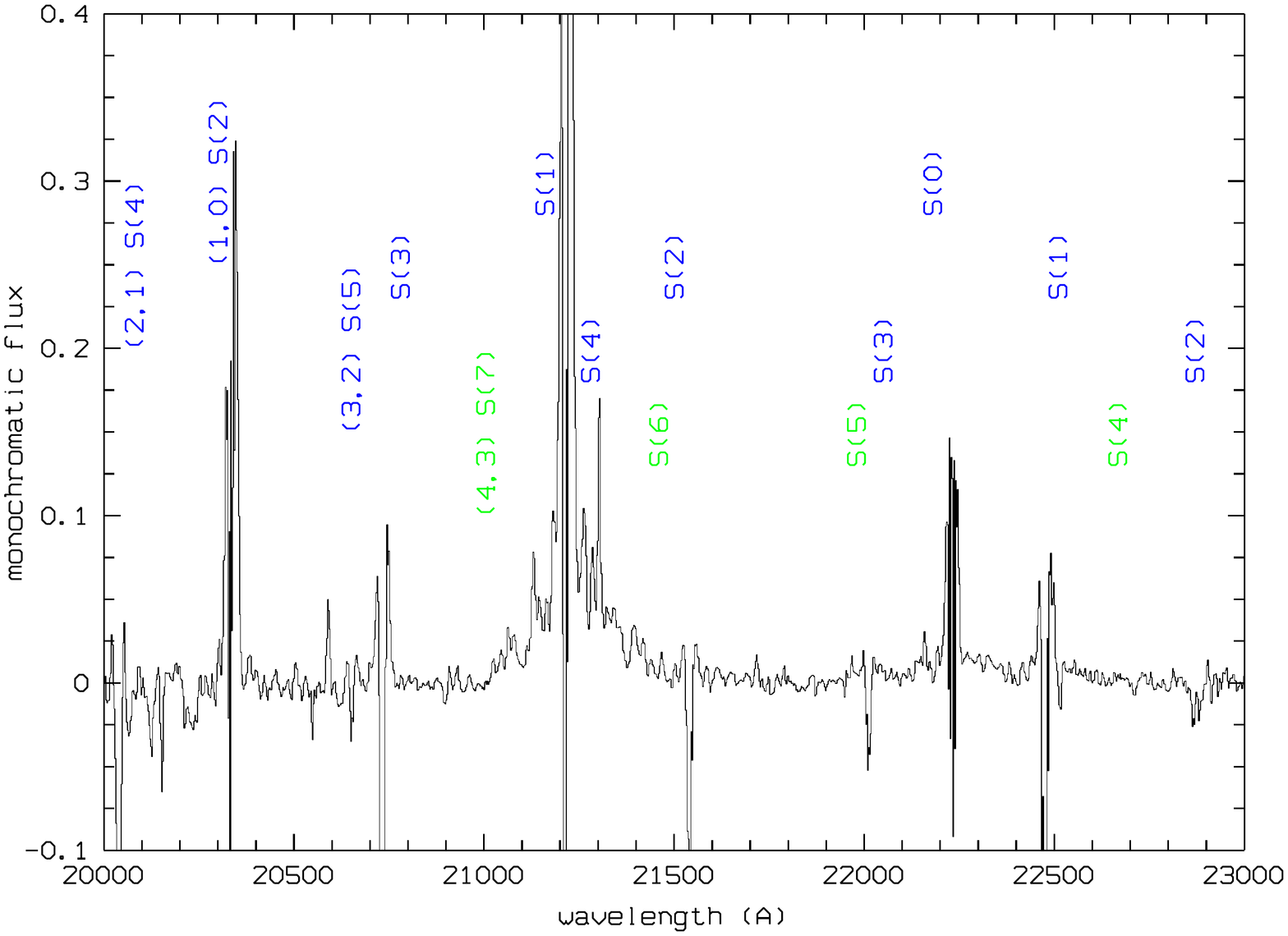}
   \caption{
Residuals between observed and modeled H$_2$ spectra as shown in Fig.~7,
covering the range of 20\,000--23\,000 \ang. The broad line wings in the (1,0) S(1)
line arise from an instrumental defect of SOFI.
               }
              \label{resKs}
    \end{figure*}

As far as the possibility is concerned that the observed H$_2$ emission arises from
molecules which reform after the passage of a fast, dissociative shock, firm statements
are more difficult to reach. The H$_2$ emissivities produced from reforming H$_2$ in diffuse and
dense gas were calculated by Tin\'e et al. (\cite{tine03}) and by LeBourlot et al.
(\cite{lebourlot02}). Tin\'e et al. (\cite{tine03}) calculated an emission spectrum 
of H$_2$ produced via an Eley-Rideal process on graphite.
In Fig.~\ref{full}
we reproduce the H$_2$ emission spectrum which results from our two-component
gas model (2750~K + 1\% 6000~K). For the sake of simplicity, the width of the H$_2$
emission lines is kept constant at 200 \ang\ over the spectral range
of 5000 \ang\ to 5 $\mu$m of Fig.~\ref{full}. A comparison with Figs.~2 and 3 of Tin\'e et al.
(2003) does not allow us to rule out the presence of re-forming H$_2$ molecules in HH91A
nor does it support the idea that reformation occurs.  
In particular, the very strong emission in the (0,0) S(9) line predicted by 
Tin\'e et al. (2003) in their mechanism is expected from thermal gas at 2750~K as 
well. The models presented by Casu \& Cecchi-Pestellini (2005) predict
very large column densities in very high rotational levels (J $>$ 20) 
of H$_2$. The wavelengths of the very high rotational lines are 
not covered by our observations.

\section{Conclusions}
The findings presented here are summarized as follows:

   \begin{enumerate}
\item From the analysis of some 200 emission lines of molecular hydrogen which are 
detected towards HH91A, it is concluded that the emission arises from thermally excited 
H$_2$, where the bulk of the gas is at a temperature of 2750~K and where 1\%
of the gas is at a temperature of 6000~K.  The total column density of the shocked H$_2$ is 
N(H$_2) = 10^{18}$ cm$^{-2}$.  \item Emission from very high-excitation lines in the 
(6,4), (6,3), (7,4), and (8,4) bands is detected, with excitation energies of the
corresponding ro-vibrational levels of up to 40\,000~K. The fluxes in these high-excitation
lines are consistent with the expectations from a thermally excited gas. 
\item It is suggested that the H$_2$ emission arises from a slow, non-dissociative J-shock.
A comparison with model calculations shows that contributions from non-thermal excitation
scenarios, such as H$_2$ pumping by Ly$\alpha$ or UV radiation, or collisional 
excitations by non-thermal, fast electrons, are not significant. 
\item The results are inconclusive as far as the presence of H$_2$ emission from
reforming molecules is concerned. 

   \end{enumerate}

\begin{acknowledgements}
The insightful and constructive comments of the referee, Chris Davis, are gratefully
acknowledged. 
The assistance of Nicolas Cardiel and Alberto Aguirre during the PMAS observations
and of Sebastian Sanchez during the reduction of the PMAS data is acknowledged.
\end{acknowledgements}

\longtab{1}{
\begin{longtable}{lrrr}
\caption{\label{tabletwo}Near-infrared line detections obtained with SOFI}\\
\hline\hline
  Line &  Wavelength & Flux $F$ & $N$(v'J') \\
                        &  (\ang)     & $(10^{-19}$Wm$^{-2})$
                                      & $(10^{14} $cm$^{-2})$\\
\hline
\endfirsthead
\caption{continued.}\\
\hline\hline
  Line &  Wavelength & Flux $F$& $N$(v'J') \\
                        &  (\ang)     & $(10^{-19}$Wm$^{-2})$
                                      & $(10^{14} $cm$^{-2})$\\
\hline
\endhead
\hline
\endfoot
grism GB order 1 \\
( 6, 3)S( 8)  & 0.953  &    4.7(  1.0 ) &  0.3 (  0.1 )\\ 
( 6, 3)S( 4)  & 0.959  &    1.7(  1.0 ) &  0.1 (  0.0 )\\
( 6, 3)S( 3)  & 0.966  &    4.1(  1.0 ) &  0.4 (  0.1 )\\ 
( 7, 4)S( 3)*  & 1.028  &    4.8(  1.0 ) &  0.3 (  0.1 )\\ 
( 7, 4)S( 5)*  & 1.029  &    4.8(  1.0 ) &  0.3 (  0.1 )\\ 
( 5, 2)O( 4)*  & 1.039  &    5.0(  1.0 ) &  6.0 (1.2) \\ 
( 6, 3)Q( 3)*  & 1.039  &    5.0(  1.0 ) &  1.4 (0.3) \\
\end{longtable}
}% End \longtab

\Online

\longtab{2}{
\begin{longtable}{lrrrrr}
\caption{\label{tablethree}Theoretical H$_2$ line fluxes}\\
\hline\hline
  Line &  Wavelength & $E_\mathrm{up}$ & $F_\mathrm{2750K}$ & $F_\mathrm{6000K}$
 & $F_\mathrm{tot}$ \\
       &  ($\mu$m)     & (cm$^{-1}$)  
& $(10^{-19}$Wm$^{-2})$
& $(10^{-19}$Wm$^{-2})$
& $(10^{-19}$Wm$^{-2})$\\
\hline
\endfirsthead
\caption{continued.}\\
\hline\hline
  Line &  Wavelength & $E_\mathrm{up}$ & $F_\mathrm{2750K}$ & $F_\mathrm{6000K}$ 
& $F_\mathrm{tot}$ \\
       &  ($\mu$m)     & (cm$^{-1}$)  
& $(10^{-19}$Wm$^{-2})$
& $(10^{-19}$Wm$^{-2})$
& $(10^{-19}$Wm$^{-2})$\\
\hline
\endhead
\hline
\endfoot

 (0,0)S( 2)  &    12.279  &    1168.8  &       2.5  &       0.0  &       2.5\\  
 (0,0)S( 3)  &     9.665  &    1740.2  &      30.8  &       0.5  &      31.3\\  
 (0,0)S( 4)  &     8.026  &    2414.8  &      27.6  &       0.5  &      28.1\\  
 (0,0)S( 5)  &     6.909  &    3187.6  &     164.8  &       4.1  &     168.8\\  
 (0,0)S( 6)  &     6.109  &    4051.7  &      87.0  &       2.7  &      89.7\\  
 (0,0)S( 7)  &     5.511  &    5002.0  &     344.7  &      14.2  &     358.9\\  
 (0,0)S( 8)  &     5.053  &    6030.8  &     130.7  &       7.2  &     137.9\\  
 (0,0)S( 9)  &     4.695  &    7132.0  &     393.4  &      29.7  &     423.1\\
\end{longtable}
}% End \longtab

\longtab{3}{
\begin{longtable}{lrrrrr}
\caption{\label{tablefour}Theoretical H$_2$ line fluxes, with modeled fluxes
between $(0.5-1) \times 10^{-19}$ Wm$^{-2}$.}\\
\hline\hline
  Line &  Wavelength & E$_\mathrm{up}$ & F$_\mathrm{2750K}$ & F$_\mathrm{6000K}$
 & F$_\mathrm{tot}$ \\
       &  ($\mu$m)     & (cm$^{-1}$)  
& $(10^{-19}$Wm$^{-2})$
& $(10^{-19}$Wm$^{-2})$
& $(10^{-19}$Wm$^{-2})$\\
\hline
\endfirsthead
\caption{continued.}\\
\hline\hline
  Line &  Wavelength & E$_\mathrm{up}$ & F$_\mathrm{2750K}$ & F$_\mathrm{6000K}$ 
& F$_\mathrm{tot}$ \\
       &  ($\mu$m)     & (cm$^{-1}$)  
& $(10^{-19}$Wm$^{-2})$
& $(10^{-19}$Wm$^{-2})$
& $(10^{-19}$Wm$^{-2})$\\
\hline
\endhead
\hline
\endfoot

 (0,0)S( 1)  &    17.035  &     705.5  &       0.9  &       0.0  &       0.9\\  
 (1,0)O(12)  &     6.308  &    9883.8  &       0.5  &       0.1  &       0.6\\  
 (1,0)O(13)  &     7.126  &   10927.1  &       0.5  &       0.1  &       0.7\\  
 (1,0)S(12)  &     1.639  &   14399.1  &       0.4  &       0.2  &       0.6\\  
 (1,0)Q(16)  &     2.891  &   16936.2  &       0.4  &       0.5  &       1.0\\  
 (1,0)S(16)  &     1.634  &   19595.7  &       0.2  &       0.5  &       0.7\\  
 (1,0)Q(19)  &     3.118  &   20957.7  &       0.1  &       0.5  &       0.6\\  
 (1,0)S(18)  &     1.659  &   22333.8  &       0.1  &       0.8  &       1.0\\
\end{longtable}
}% End \longtab

\end{document}